\documentclass[
aip,
reprint,%
]{revtex4-1}

\usepackage{graphicx}
\usepackage{dcolumn}
\usepackage{bm}

\usepackage[utf8]{inputenc}
\usepackage[T1]{fontenc}
\usepackage{mathptmx}
\usepackage{etoolbox}
\usepackage{color}
\usepackage{siunitx}
\usepackage{amsmath}
\usepackage{amssymb}
\usepackage{bm}

\newcommand{\diff}{\,\mathrm{d}}
\newcommand{\pd}[2]{\frac{\partial #1}{\partial #2}}

\newcommand{\bfe}{\mathbf{e}}
\newcommand{\bfex}{\mathbf{e}_x}
\newcommand{\bfey}{\mathbf{e}_y}

\newcommand{\bfr}{\mathbf{r}}
\newcommand{\dbfr}{\,\mathrm{d}\mathbf{r}}
\newcommand{\bfk}{\mathbf{k}}

\newcommand{\ltf}{l_{\rm TF}}
\newcommand{\ktf}{k_{\rm TF}}
\newcommand{\leff}{L_{\rm eff}}

\newcommand{\varepssol}{\varepsilon_{\rm s}}

\definecolor{darkblue}{rgb}{0,0,0.6}
\definecolor{darkred}{rgb}{0.6,0,0}
\definecolor{forestgreen}{rgb}{0.13,0.55,0.13}
\usepackage[colorlinks=true,urlcolor=darkblue,citecolor=darkblue,linkcolor=darkred]{hyperref}

\begin{document}



\title{Brownian dynamics simulations of electric double-layer capacitors with tunable metallicity}



\author{Paul Desmarchelier}
\affiliation{Sorbonne Universit\'e, CNRS, Laboratoire PHENIX (Physicochimie des Electrolytes et Nanosyst\`emes Interfaciaux), 4 place Jussieu, 75005 Paris, France}

\author{Alexandre P. dos Santos}
\affiliation{Instituto de Física, Universidade Federal do Rio Grande do Sul, Caixa Postal 15051, CEP 91501-970 Porto Alegre, RS, Brazil}

\author{Yan Levin}
\affiliation{Instituto de Física, Universidade Federal do Rio Grande do Sul, Caixa Postal 15051, CEP 91501-970 Porto Alegre, RS, Brazil}

\author{Benjamin Rotenberg}
\email[]{benjamin.rotenberg@sorbonne-universite.fr}
\affiliation{Sorbonne Universit\'e, CNRS, Laboratoire PHENIX (Physicochimie des Electrolytes et Nanosyst\`emes Interfaciaux), 4 place Jussieu, 75005 Paris, France}
\affiliation{Réseau sur le Stockage Electrochimique de l’Energie (RS2E), FR CNRS 3459, 80039 Amiens Cedex, France}


\date{\today}

\begin{abstract}
We introduce an efficient description of electrodes, characterized by their Thomas-Fermi screening length $\ltf$ inside the metal, for Brownian dynamics (BD) simulations of capacitors. Within a Born-Oppenheimer approximation for the electron charge density inside the electrodes, we derive the effective many-body potential for ions in an implicit solvent between Thomas-Fermi electrodes, taking into account the constraints of applied voltage and of global electro-neutrality of the system, as well as the 2D periodic boundary conditions along the electrode surfaces. We derive the average charge and the fluctuation-dissipation relation for the differential capacitance, highlighting the contribution of the fluctuations of the net ionic dipole moment, as well as those from the solvent polarization and of the electron density, whose fluctuations are suppressed within the Born-Oppenheimer description. We demonstrate the relevance of this model by validating its predictions against known results for the force on ions as a function of the ion-surface distance in simple geometries. The equilibrium ionic density profiles from BD simulations are in excellent agreement with those from an explicit electrode model for perfect metals, and are obtained at a significantly lower computational cost.  Finally, we discuss with the present model the effect of the Thomas-Fermi screening length on the equilibrium ionic density profiles and the capacitance. While limited to parallel plate capacitors, the present simulation method allows to consider larger systems, lower concentrations, and longer time scales concentrations than molecular simulations in order to predict the electrochemical properties of Thomas-Fermi capacitors and correlate them with the ion dynamics. 
\end{abstract}

\pacs{}

\maketitle 

\section{Introduction}

The interface between metals and liquid electrolytes plays a key role in electrochemical energy storage devices (batteries, electric double layer capacitors), electrocatalysis, or electrochemical sensing. In these contexts, the interfacial properties result from the mutual influence of the  electron density inside the metal and of the charge distribution due to the solvent molecules and the ions inside the liquid. This interplay controls in particular the equilibrium properties such as the ionic densities and electrostatic potential profiles, or the capacitance that links the charge accumulated on both sides of the interface to the potential drop across it. From the theoretical point of view, the predictions of equilibrium properties generally rely on the mean-field Poisson-Boltzmann theory, its linearized limit (Debye-H\"uckel theory), or its extensions~\cite{parsons_electrical_1990, kornyshev_double-layer_2007, goodwin_mean-field_2017, baskin_improving_2017, mazur_understanding_2024}. More advanced liquid state theories such as integral equations, field theory or classical Density Functional Theory (cDFT) have been developed to include in particular electrostatic correlations~\cite{kjellander_correlation_1984, zhou_image_2024}, the finite size of the ions (\textit{e.g.} within the primitive model of electrolytes consisting of charged hard spheres)~\cite{cats_primitive_2021, cats_in-plane_2023}, the explicit polarization of the solvent in addition to the ionic densities~\cite{levy_dielectric_2012, levy_dipolar_2013, bruch_incorporating_2024, bruch_variational_2025} or even molecular features of the solvent around the ions~\cite{jeanmairet_molecular_2013, jeanmairet_study_2019}. For dynamical properties, most results have been obtained at the same level of description as Poisson-Boltzmann theory (point ions in an implicit solvent interacting via the mean-field electrostatic potential), within Poisson-Nernst-Planck theory. This model allowed in particular to investigate the charging dynamics in capacitors~\cite{bazant_diffuse_2004, janssen_transient_2018, palaia_charging_2025, palaia_poisson-nernst-planck_2025} or their frequency-dependent impedance~\cite{barbero_theory_2008,antonova_ambipolar_2020}. Here again, progress \textit{e.g.} with time-dependent cDFT allowed to analyze the effect of some physical ingredients such as the finite size of the ions or electrokinetic couplings on the dynamical response of capacitors~\cite{kilic_steric_2007, jiang_time-dependent_2014, asta_lattice_2019, ma_dynamic_2022}. 

In the last decades, these theoretical predictions were complemented by simulations at various levels of description. Ab initio or hybrid QM/MM molecular dynamics (MD) simulations provide the most accurate description but entail a large computational cost that limits the system size and trajectory length that can be simulated~\cite{bonnet_first-principle_2012, elliott_qmmd_2020, le_modeling_2021, takahashi_accelerated_2022, gross_ab_2022, sakong_structure_2022, gross_challenges_2023, grisafi_predicting_2023}. Classical all-atom and coarse-grained MD simulations offer a good trade-off between accuracy of the description and computational cost, and the implementation in generic or dedicated open source simulation packages~\cite{ weik_espresso_2019, nguyen_incorporating_2019,marin-lafleche_metalwalls_2020, coretti_metalwalls_2022, ahrens-iwers_electrode_2022} of methods to sample systems in a statistical ensemble where the voltage between the electrodes is imposed~\cite{siepmann_influence_1995, reed_electrochemical_2007} have enabled their use for a variety of electrode/electrolyte interfaces.  We refer the reader \emph{e.g.} to Refs.~\citenum{scalfi_microscopic_2021} and~\citenum{jeanmairet_microscopic_2022} for recent reviews on these topics. Such simulations allow in particular to compute the capacitance or the frequency-dependent impedance of electrochemical cells and to correlate these observables with the microscopic properties of the confined liquid~\cite{limmer_charge_2013, scalfi_charge_2020, pireddu_frequency-dependent_2023, pireddu_impedance_2024}.

In order to further decrease the computational cost of particle based simulations, implicit solvent models, where the effect of the solvent on ion-ion interactions is captured by the solvent permittivity and the ions evolve according to Langevin or Brownian dynamics (BD), instead of Newton's equation of motion. The effects of the confining dielectric walls or perfect conductors and of voltage are then introduced as external potentials or enter in the ion-ion interactions, in order to model driven electrolytes confined in slit pores or capacitors under an applied voltage~\cite{arnold_electrostatics_2002, tyagi_iterative_2010, breitsprecher_electrode_2015,  nguyen_incorporating_2019, maxian_fast_2021, jimenez-angeles_surface_2023, dos_santos_modulation_2023, pogharian_electric_2024}. For perfect conductors, efficient simulations at this level of description, taking into account periodic boundary conditions inherent to particle-based simulations in condensed matter systems, have been proposed within the Green's function formalism to compute effective ion-ion interactions~\cite{ dos_santos_simulations_2017, girotto_simulations_2017, malossi_simulations_2020, telles_efficient_2024}. An alternative approach for BD simulations, combining the description of explicit constant-potential electrodes using in MD simulations and ions with an implicit solvent described by its permittivity has also been proposed by Cats \textit{et al.}~\cite{cats_capacitance_2022}. Dynamical properties, such as the electric double layer relaxation after a charge transfer event~\cite{grun_relaxation_2004}, the charge fluctuations of an electrode due to the interfacial ion dynamics and redox reaction~\cite{katelhon_simulation-based_2012, krause_brownian_2014}, or the frequency- and field-dependent response of confined electrolytes~\cite{hoang_ngoc_minh_frequency_2023} have also been investigated using BD simulations.

Even though most theoretical and simulation studies to date have been limited to perfect metals, the importance of charge screening inside the metal has long been recognized. At the fundamental level, this phenomenon arises from the kinetic energy of the electrons, which has to be described using quantum mechanics. Thomas-Fermi (TF) theory~\cite{thomas_calculation_1927, fermi_metodo_1927} introduces this within a simple kinetic energy functional. Within a linearization approximation, the electrostatic potential then satisfies an equation identical to the Debye-H\"uckel equation for electrolytes, with the screening length replaced by the so-called Thomas-Fermi screening length, $\ltf$, that depends on the density of states of the metal. This behavior can also be understood in terms of non-local electrostatics, via a wave vector dependent permittivity that depends on $\ltf$. TF theory has been used to investigate the consequence of such screening inside the metal on ion-wall, ion-ion interactions or interfacial capacitance~\cite{kornyshev_image_1977, vorotyntsev_electrostatic_1980, rochester_interionic_2013, lee_quantum_2016, lee_ionimage_2016, hedley_what_2025} (note that the electrostatic problem is the same as the one for an ion in a slit pore surrounded by an electrolyte treated at the Debye-H\"uckel level, that was solved analytically in Ref.~\citenum{levin_electrostatics_2006}). The consequences on the collective interfacial behavior has been demonstrated experimentally, for example on the capillary freezing of ionic liquids confined depending on the metallicity of the substrate~\cite{comtet_nanoscale_2017}, which could be rationalized within TF theory~\cite{kaiser_electrostatic_2017}, or the kinetics of electron transfer at twisted graphene bilayers~\cite{yu_tunable_2022}. Several methods have also been proposed to introduced TF theory in atomistic simulations, to examine the consequences of screening inside the metal on the interfacial properties~\cite{scalfi_charge_2020, schlaich_electronic_2022, goloviznina_accounting_2024, nair_induced_2025, nair_ions_2025, coello_escalante_microscopic_2024}. They suffer however from the above-mentioned limitations in terms of system size, salt concentration and trajectory length due to the computational cost of molecular simulations.

In the present work, we introduce an efficient description of electrodes, characterized by their Thomas-Fermi screening length $\ltf$ inside the metal, for Brownian dynamics (BD) simulations of capacitors. In Section~\ref{sec:theory}, we derive within a Born-Oppenheimer approximation for the electron charge density inside the electrodes, the effective many-body potential for ions in an implicit solvent between Thomas-Fermi electrodes, taking into account the constraints of applied voltage and of global electro-neutrality of the system, as well as the 2D periodic boundary conditions along the electrode surfaces. We derive the average charge and the fluctuation-dissipation relation for the differential capacitance, highlighting the contribution of the fluctuation of the net ionic dipole moment, as well as those from the solvent polarization and of the electron density, whose fluctuations are suppressed within the Born-Oppenheimer description. In section~\ref{sec:results}, we demonstrate the relevance of this model by validating its predictions against know results for the force on ions as a function of the ion-surface distance in simple geometries. The equilibrium ionic density profiles from BD simulations are in excellent agreement with those from an explicit electrode model for perfect metals, and are obtained at a significantly lower computational cost. Finally, we discuss with the present model the effect of the Thomas-Fermi screening length on the equilibrium ionic density profiles and the capacitance. While limited to parallel plate capacitors, the present simulation method allows to consider larger systems, lower concentrations, and longer time scales concentrations than molecular simulations in order to predict the electrochemical properties of Thomas-Fermi capacitors and correlate them with the ion dynamics.

\section{Theory}
\label{sec:theory}

\subsection{System}
\label{sec:theory:system}

\begin{figure}[ht!]
    \begin{center}
        \includegraphics[width=0.45\textwidth]{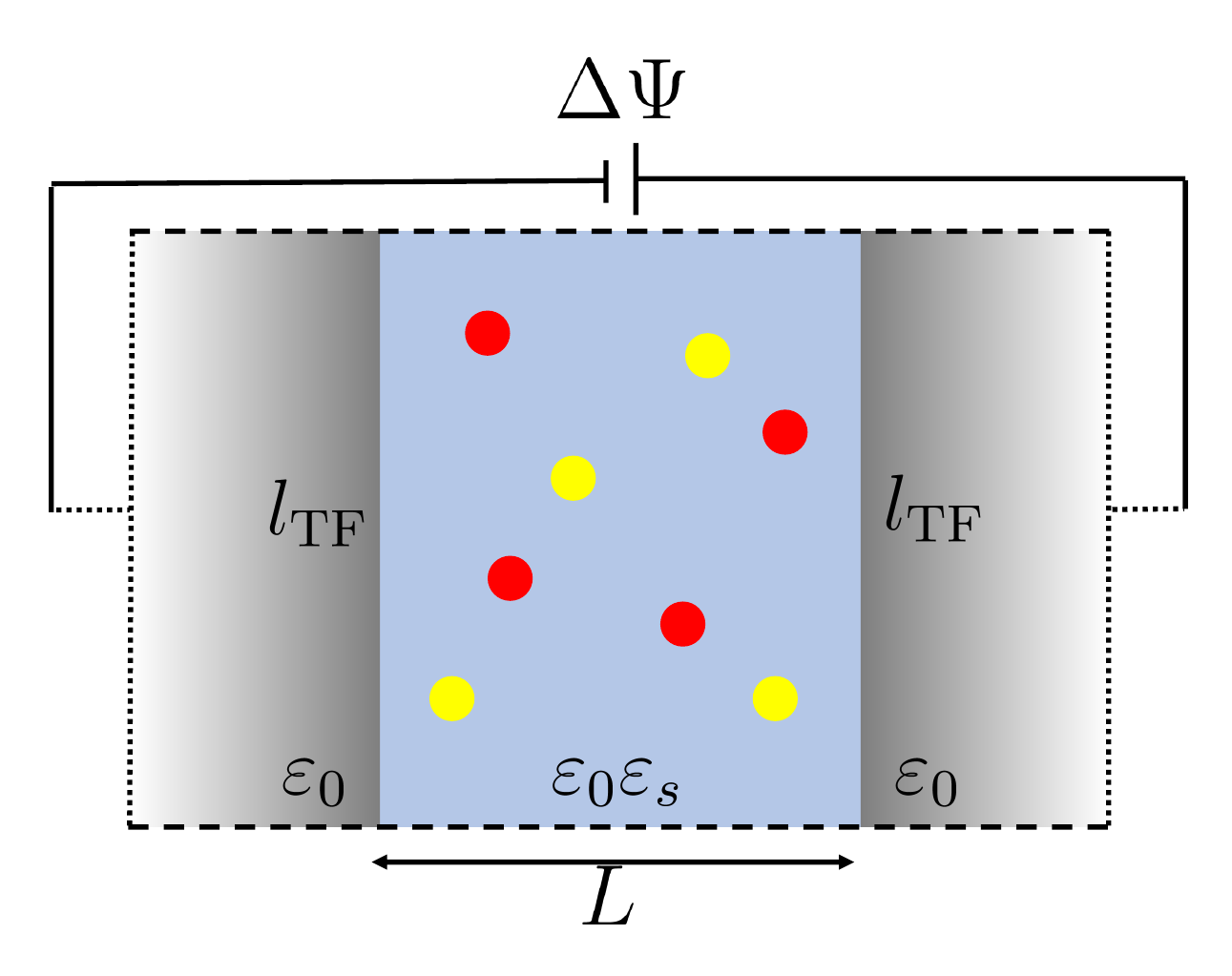} 
    \end{center}
    \caption{
    Capacitor consisting of two Thomas-Fermi electrodes separated over a distance $L$ by an electrolyte solution, under an applied voltage $\Delta\Psi$. The electrolyte consists of explicit ions in an implicit solvent with permittivity $\varepsilon_0\varepssol$, while the electrodes are characterized by a permittivity $\varepsilon_0$ and a Thomas-Fermi screening-length $\ltf$. Periodic boundary conditions in the $x$ and $y$ directions along the electrode-electrolyte interfaces, with box dimensions $L_x$ and $L_y$, are indicated by dashed lines. The dotted lines indicate the connexion with an electric circuit for $z\to\pm\infty$ that imposes the external voltage.
    }
    \label{fig:system}
\end{figure}

We consider a capacitor consisting of two electrodes labeled left, $l$, and right, $r$,  separated by an electrolyte solution ($s$), as illustrated in Fig.~\ref{fig:system}. These regions are characterized by their position $z<-L/2$, $z>L/2$ and $z\in[-L/2,L/2]$, respectively, while the system is periodic in the $x$ and $y$ directions. Note that the reference to left and right is only for convenience and is not related to a specific orientation of the capacitor. The electrodes are modeled by a continuous background of positive charges, with charge density $\rho^{\rm back}_{\rm l,r}(\bfr)= +e n_{\rm l,r}^0$, with $n_{\rm l,r}^0$ the average electron density in each electrode, and continuous electron distributions, with charge density $\rho^e_{\rm l,r}(\bfr)=-e [n_{\rm l,r}^0+\delta n_{\rm l,r}(\bfr)]$, while the electrolyte is modeled by point charges $q_i$ at positions $\bfr_i$ in a solvent with permittivity $\varepsilon_0\varepssol$. 
The corresponding charge densities are
\begin{align}
	\rho_{\rm l,r}(\bfr)&= \rho^{\rm back}_{\rm l,r}(\bfr) + \rho^e_{\rm l,r}(\bfr) = -e\delta n_{\rm l,r}(\bfr)\\
	\rho_{\rm ions}(\bfr) &=\sum_{i=1}^{N_{\rm ions}}q_i\delta(\bfr-\bfr_i)
\end{align}
The total charge density can be written as $\rho^{\rm tot}(\bfr) = \rho_{\rm l}(\bfr) + \rho_{\rm ions}(\bfr) + \rho_{\rm r}(\bfr)$, even though only one of the three terms contributes in each region. It is the source of an electrostatic potential $\phi(\bfr)$ satisfying the Poisson equation $\Delta\phi=-\rho^{\rm tot}/\varepsilon_0\varepsilon$ in each region of space ($\varepsilon=1$ inside the electrodes and $\varepssol$ in the electrolyte), with $\Delta$ the Laplacian operator, as well as continuity of $\phi$ and $\varepsilon\phi'$ at the interfaces between regions. 
The values of the potentials for $z\to\pm\infty$ will be discussed in the following. Since both ions and electrons are part of the system, the total electrostatic energy is
\begin{align}
\label{eq:UCoul}
U_{\rm Coul}[n_{\rm l},n_{\rm r},\rho_{\rm ions}] & = \frac{1}{2}\int\rho(\bfr)\phi(\bfr) \dbfr \nonumber \\
& \hspace{-1.5cm} = \frac{1}{2}\int_{\rm l}\rho_{\rm l}\phi \dbfr + \frac{1}{2}\int_{\rm s}\rho_{\rm ions}\phi\dbfr+\frac{1}{2}\int_{\rm r}\rho_{\rm r}\phi \dbfr  \; ,
\end{align}
where in the second line we have separated the integral over the three relevant regions of space. The electrostatic potential can be written as a $\phi(\bfr)=\int \rho^{\rm tot}(\bfr')G(\bfr,\bfr') \diff \bfr'$, where the Green's function $G(\bfr,\bfr')$ describes the electrostatic interaction between unit point charges at $\bfr$ and $\bfr'$ and the integral runs over the whole space occupied by the electrodes and the electrolyte. This highlights the quadratic form of the electrostatic energy and the fact that a factor of $1/2$ is needed to avoid double-counting. Note that $G(\bfr,\bfr')$ diverges when $\bfr'\to\bfr$, for example as $1/4\pi\varepsilon_0\varepssol|\bfr'-\bfr|$ in the electrolyte solution (see Section~\ref{sec:theory:BO:EulerLagrangeecPot} and Appendix~\ref{sec:appendix:phi} for more details). This diverging self-interaction term should be removed when computing the energy.

In addition to the Coulomb energy, we consider the kinetic energy of the electrons described by
\begin{align}
\label{eq:Texact}
T[n_{\rm l},n_{\rm r}] &= \int_{\rm l}\tau(n_{\rm l}(\bfr))\dbfr + \int_{\rm r}\tau(n_{\rm r}(\bfr))\dbfr
\end{align}
where
\begin{align}
\label{eq:tau}
\tau(n) &=A n^{5/3} = \frac{3h^2}{40m_e}\left(\frac{3}{\pi}\right)^{2/3} n^{5/3} \; ,
\end{align}
with $h$ Planck's constant and $m_e$ the mass of the electron,
the Thomas-Fermi (TF) kinetic energy functional. This term, which is of quantum-mechanical origin, results together with the Coulomb energy in screened electrostatic interactions that can be described within the classical Thomas-Fermi model.

While the system is canonical for the electrolyte (fixed number of ions $N_{\rm ions}$) and for the background positive charges in the electrodes, it is grand-canonical for the electrons: the electron densities in both electrodes fluctuate by exchanging electrons with reservoirs setting their electrochemical potentials $\mu_{\rm l,r}$ (one value per electrode). In addition, we impose the global electroneutrality of the system:
\begin{align}
\label{eq:neutrality}
 \int_{\rm l} (-e)\delta n_{\rm l} \dbfr  +  \int_{\rm s} \rho_{\rm ions}  \dbfr + \int_{\rm r} (-e)\delta n_{\rm r} \dbfr&=0 \; .
\end{align}
In practice, this amounts to imposing that electrons are exchanged between the electrodes, rather than with two independent reservoir, hence (see below) to imposing and a finite electrochemical potential difference $\Delta\mu=\mu_{r}-\mu_{l}$ or voltage $\Delta\Psi$ between the electrodes instead of independent electrochemical potentials. Assuming that the electrolyte is also overall neutral, this implies that the total charge of the two electrodes are opposite to each other. The corresponding thermodynamic potential for the electrons is:
\begin{align}
\label{eq:Omega}
	\Omega[n_{\rm l},n_{\rm r}|\rho_{\rm ions},\mu_{\rm l},\mu_{\rm r}] = & U_{\rm Coul}[n_{\rm l},n_{\rm r},\rho_{\rm ions}] + T[n_{\rm l},n_{\rm r}] \nonumber \\
	& \hspace{-2.5cm} -\mu_{\rm l}\int_{\rm l} (n_{\rm l}^0+\delta n_{\rm l})\dbfr -\mu_{\rm r}\int_{\rm r} (n_{\rm r}^0+\delta n_{\rm r}) \dbfr\nonumber \\
	& \hspace{-2.5cm} -\lambda\left[ \int_{\rm l} (-e)\delta n_{\rm l} \dbfr +  \int_{\rm s} \rho_{\rm ions} \dbfr  +\int_{\rm r} (-e)\delta n_{\rm r} \dbfr \right]
\end{align}
where $\lambda$ is a Lagrange multiplier enforcing global electroneutrality, and the total Coulomb energy and the kinetic energy of the electrons are given by Eqs.~\ref{eq:UCoul} and~\ref{eq:Texact}, respectively. Note that we have made explicit here that we consider a functional of the electron densities $n_{\rm l,r}(\bfr)$ that depends parametrically on the ionic distribution $\rho_{\rm ions}(\bfr)$ and the electrochemical potentials $\mu_{\rm l,r}$.

\subsection{Born-Oppenheimer description}
\label{sec:theory:BO}

For a given configuration of the ions and corresponding charge distribution, $\rho_{\rm ions}(\bfr)$, we look for the electron density that minimizes the total energy under the constraints of fixed electrochemical potentials $\mu_{\rm l,r}$ and global electroneutrality. This is achieved by minimizing the thermodynamic potential Eq.~\ref{eq:Omega} with respect to the charge densities $n_{\rm l,r}$ in the electrodes, \textit{i.e.} solving the Euler-Lagrange equations:
\begin{align}
	\left.\frac{\delta \Omega[n_{\rm l},n_{\rm r}|\rho_{\rm ions},\mu_{\rm l},\mu_{\rm r}]}{\delta n_{\rm l,r}(\bfr)}\right|_{n_{\rm l,r}^{\rm BO}} &= 0 \; ,
	\label{eq:EL0}
\end{align}
where the BO superscript refers to the Born-Oppenheimer distribution of the electrons satisfying these constraints. In the following, we will not always include the BO superscript explicitly but one should remember that the equations below hold only for $n_{\rm l,r}^{\rm BO}$. Once these distributions are determined, one can define an effective (many-body) potential of the system consisting only of ions as:
\begin{align}
	V^{\rm eff}[\{\bfr_i\}|\mu_{\rm l},\mu_{\rm r}] &\equiv\Omega[n_{\rm l}^{\rm BO},n_{\rm r}^{\rm BO}|\rho_{\rm ions},\mu_{\rm l},\mu_{\rm r}] 
	\label{eq:Fdef}
\end{align}
where we have again highlighted the parametric dependence on the electrochemical potentials $\mu_{\rm l,r}$. The corresponding force acting on the ions can then be computed as the gradient with respect to the ionic positions
\begin{align}
	{\bf F}_i^{\rm eff} = -\nabla_{\bfr_i} V^{\rm eff}[\{\bfr_i\}|\mu_{\rm l},\mu_{\rm r}]
	\label{eq:Feff}
\end{align}
which accounts for the reorganization of the charge distribution within the electrodes as the ions move. This is analogous to what is done in \textit{ab initio} molecular dynamics, via the Hellmann-Feynman theorem, or to the case of explicit colloids with small ions treated at the DFT level, see Ref.~\citenum{lowen_ab_1992}.

\subsubsection{Euler-Lagrange equations}
\label{sec:theory:BO:EulerLagrange}

In order to proceed further, we assume that $|\delta n_{\rm l,r}|\ll n_{\rm l,r}^0$ and approximate the kinetic energy Eq.~\ref{eq:Texact}, up to quadratic order in the electrode charge densities for consistency with the Coulomb term, using
\begin{align}
\label{eq:Tapprox}
    \tau(n_{\rm l,r}) &=
	\tau(n^0_{\rm l,r}) + \mu^T_{\rm l,r} \delta n_{\rm l,r} + \frac{1}{2}\alpha_{\rm l,r}\delta n_{\rm l,r}^2 +\mathcal{O}(\delta n_{\rm l,r}^3)
\end{align}
where (see Eq.~\ref{eq:tau})
\begin{align}
\label{eq:muT}
\mu^T_{\rm l,r} &= \left.\pd{\tau}{n}\right|_{n_{\rm l,r}^0} 
    = \frac{5A}{3}(n_{\rm l,r}^0)^{2/3}
\\
\label{eq:alpha}
\alpha_{\rm l,r} & = \left.\pd{^2\tau}{n^2}\right|_{n_{\rm l,r}^0} 
= \frac{10A}{9}(n_{\rm l,r}^0)^{-1/3}
\; .
\end{align}
The corresponding local chemical potentials are
\begin{align}
	\frac{\delta T[n_{\rm l},n_{\rm r}]}{\delta n_{\rm l,r}(\bfr)}
	&= \mu^T_{\rm l,r} + \alpha_{\rm l,r}\delta n_{\rm l,r}(\bfr) +\mathcal{O}(\delta n_{\rm l,r}^2)
\end{align}

Minimizing the thermodynamic potential with respect to the electronic densities in the left and right electrodes yields (using Eq.~\ref{eq:Tapprox} and the fact that the Coulomb energy is quadratic in the densities) :
\begin{align}
	\frac{\delta \Omega[n_{\rm l},n_{\rm r}|\rho_{\rm ions},\mu_{\rm l},\mu_{\rm r}]}{\delta n_{\rm l,r}(\bfr)} &= 0 
    \nonumber \\
    & \hspace{-2cm }= \mu_{\rm l,r}^T+\alpha_{\rm l,r}\delta n_{\rm l,r}(\bfr)-e\phi(\bfr)-\mu_{\rm l,r} + \lambda e  \; .
	\label{eq:EL1}
\end{align}
In particular, in the bulk of the electrodes $\delta n_{\rm l,r}(\bfr)=0$ so that
\begin{align}
	\label{eq:mumuTPsi}
    \mu_{\rm l,r} &= \mu_{\rm l,r}^T -e (\Psi_{\rm l,r} - \lambda) 
\end{align}
where $\Psi_{l}=\lim_{z\to-\infty}\phi$ and $\Psi_{r}=\lim_{z\to+\infty}\phi$ are such that $\Delta\Psi=\Psi_{\rm r}-\Psi_{\rm l}$ is the voltage imposed between the two electrodes, and $\lambda$ amounts to a constant potential shift of both potentials, which does not change the voltage, as expected. As anticipated in Section~\ref{sec:theory:system}, Eq.~\ref{eq:mumuTPsi} also highlights the fact that in practice, one imposes the the difference $\Delta\mu=\mu_{r}-\mu_{l}$ rather than the two values independently. The Euler-Lagrange equations can be rewritten as
\begin{align}
	\alpha_{\rm l,r}\delta n_{\rm l,r}(\bfr) &= e \left[\phi(\bfr)-\Psi_{\rm l,r}\right] \; .
	\label{eq:EL2}
\end{align}
Since inside the electrode, the charge densities $\delta n_{\rm l,r}$ are also related to the Poisson equation $\Delta \phi = - (-e)\delta n/\varepsilon_0$, the potential satisfies the Thomas-Fermi equation in each electrode:
\begin{align}
    \Delta \phi 
    &= k_{TF}^2 \left[\phi(\bfr)-\Psi_{\rm l,r}\right]
	\label{eq:ThomasFermi}
\end{align}
where we have introduced the inverse of the Thomas-Fermi screening length (possibly different in both electrodes, even though here we consider identical electrodes)
\begin{align}
    \ktf^2 = \frac{1}{\ltf^2} = \frac{e^2}{\varepsilon_0\alpha_{\rm l,r}}
	\label{eq:defktf}
\end{align}
where $\alpha_{\rm l,r}$ is given in Eq.~\ref{eq:alpha}. Compared to the familiar form of the TF equation, usually written for a single electrode, Eq.~\ref{eq:ThomasFermi} differs in the constant potential shifts $\Psi_{\rm l,r}$ in each electrode that enforce the applied voltage between electrodes. As shown below  (see Section~\ref{sec:theory:BO:Veff}), the constant potential shift $\lambda$ does not enter in the effective many-body potential between the ions, and the electrode potentials only appear as the voltage $\Delta\Psi$.

\subsubsection{Electrostatic potential}
\label{sec:theory:BO:EulerLagrangeecPot}

The electrostatic potential in the electrodes satisfies the above Thomas-Fermi equations Eq.~\ref{eq:ThomasFermi}, together with the boundary conditions $\Psi_{l}=\lim_{z\to-\infty}\phi$ and $\Psi_{r}=\lim_{z\to+\infty}\phi$ such that $\Delta\Psi=\Psi_{\rm r}-\Psi_{\rm l}$. In the electrolyte, it satisfies the Poisson equation $\Delta \phi = - \rho_{\rm ions}^{tot} / \varepsilon_0\varepssol$ with $\varepssol$ the solvent permittivity. At the electrode/electrolyte interfaces, $\phi$ is continuous and its derivative is such that $\varepsilon\phi'$ is continuous. 

The full electrostatic problem is conveniently expressed as the sum of two contributions $\phi=\phi_{\Delta\Psi}+\phi_{\rm ions}$, where $\phi_{\Delta\Psi}$ is the solution in the presence of the voltage $\Delta\Psi$ but in the absence of ions, while $\phi_{\rm ions}$ is the solution in the presence of ions and in the absence of voltage. The former only depends on the $z$ coordinate:
\begin{eqnarray}
\label{eq:phi:dpsi}
\phi_{\Delta\Psi}(z)=
	\begin{cases} 
		\displaystyle \Psi_{l}+\Delta\Psi \, \frac{\varepssol\ltf}{\leff}e^{\frac{1}{2}k_{TF}(L+2z)}  & -\infty<z<-\frac{L}{2}  \\
		\displaystyle \frac{\Psi_{l}+\Psi_{r}}{2}+\Delta\Psi\frac{z}{\leff}	& -\frac{L}{2}<z<\frac{L}{2} \\
		\displaystyle\Psi_{r} -  \Delta\Psi\frac{\varepssol\ltf}{\leff}e^{\frac{1}{2}k_{TF}(L-2z)}	& \frac{L}{2} <z<+\infty\\
	\end{cases}
\end{eqnarray}
where we have introduced the effective length (further discussed in Section~\ref{sec:theory:BO:Charge}):
\begin{align}
\label{eq:Leff}
    \leff &=L+2\varepssol\ltf \; .
\end{align}

The potential due to the ions can further be separated into their individual contributions as $\phi_{\rm ions}=\sum_{i}^{N_{\rm ions}}\phi_i$, where each term is the potential satisfying the problem where the ion distribution is limited to a single ion with charge $q_i$ at position $\bfr_i$ and its periodic images, in the absence of voltage:
\begin{align} 
\label{eq:rhoi}
	\rho_{i}(\mathbf{r}) & =\sum^{\infty}_{m_x=-\infty} \,\sum^{\infty}_{m_y=-\infty} q_i \,\delta(\mathbf{r}-\mathbf{r}_i+m_xL_x \bfe{x}+m_yL_y\bfey).
\end{align}
Following Ref.~\citenum{dos_santos_simulations_2017} for the case of a perfect metal ($\ltf=0$), we solve the problem in reciprocal space in the $x$ and $y$ directions by looking for a Green's function of the form
\begin{align} \label{eq:phi:phii} 
	\phi_{i}(\mathbf{r}) =\frac{q_i}{L_xL_y}\sum_{\mathbf{k}}g_{\mathbf{k}}(z_i,z)\cos\left[\mathbf{k}\cdot(\mathbf{r}-\mathbf{r}_i)\right],
\end{align}
with $\mathbf{k}=(\frac{2\pi m_x}{L_x},\frac{2\pi m_y}{L_y})$. This leads to solutions of the form $g_{{k}} = A e^{-kz}+B e^{kz}$ with $k=|\mathbf{k}|$ between the electrodes and $g_{{k}}=C_\pm e^{\pm\chi_{TF} z}$ where
\begin{align}
    \label{eq:defchiTF}
    \chi_{TF}=\sqrt{k^2+k_{TF}^2}
\end{align}inside the electrodes, the constants being determined by the boundary conditions. In addition to the one mentioned above-mentioned for $z\to\pm\infty$ and at $z=\pm L/2$, one also needs to account for continuity of $g_{\mathbf{k}}(z_i,z)$ at $z=z_i$ and the discontinuity of its derivative:
\begin{align}
\varepssol \left.\pd{g_{\mathbf{k}}(z_i,z)}{z}\right|_{z=z_i^+}=&\displaystyle\varepssol \left.\pd{g_{\mathbf{k}}(z_i,z)}{z}\right|_{z=z_i^-} - \frac{q_i}{\epsilon_0}
\end{align}
The complete solutions are provided in Appendix~\ref{sec:appendix:phi}. It can be checked in particular that $\phi_{i}(\mathbf{r})$ diverges as $1/4\pi\varepsilon_0\varepssol|\bfr-\bfr_i|$ for $\bfr\to\bfr_i$, as expected. As already mentioned in Section~\ref{sec:theory:system}, this diverging self-interaction term should be removed when computing the energy.

\subsubsection{Electrode charge}
\label{sec:theory:BO:Charge}

The electrostatic potential also allows to express the local charge induced inside each electrode via the TF Eq.~\ref{eq:ThomasFermi} and the Poisson equation. In turn, this provides the total charge of the electrodes $Q_{\rm l,r} =(-e)\int_{\rm l,r}\delta n_{\rm l,r}(\bfr)\dbfr$ induced by the applied voltage and the presence of ions. When the electrolyte is neutral, the electrodes are oppositely charged and we can introduce the charge of the right electrode $Q=Q_{\rm r}=-Q_{\rm l}$. Using Eqs.~\ref{eq:phi:dpsi} and~\ref{eq:phi:phiions} the electrode charge can be expressed as:
\begin{align}
\label{eq:totalcharge}
Q[\{\bfr_i\}|\Delta\Psi] &= L_xL_y\frac{\varepsilon_0\varepssol k_{TF}\Delta\Psi}{(2\varepssol +k_{TF}L)} 	-
		\frac{k_{TF}}{(2\varepssol +k_{TF}L)}\sum_{i}^{N_{\rm ions}}q_iz_i
		\nonumber \\
        & = C_{0}\Delta\Psi - \frac{M_{\rm ions}}{\leff}
\end{align} 
where we have introduced 
\begin{align}
\label{eq:Mions}
M_{\rm ions}=\sum_{i}q_iz_i
\end{align}
the dipole (along the $z$ direction) of the ion distribution and
the capacitance of the ion-free capacitor such that
\begin{align}
\frac{L_xL_y}{C_0} = \frac{L}{\varepsilon_0\varepssol}+\frac{2\ltf}{\varepsilon_0}
=\frac{\leff}{\varepsilon_0\varepssol}
\label{eq:capaempty}
\end{align}
and we also used the electroneutrality of the electrolyte to introduce $\phi_{\Delta\Psi}$ in the second equality. Note that Eq.~\ref{eq:capaempty} can be interpreted as an equivalent circuit, with two capacitors each with capacitance per unit area $\varepsilon_0/\ltf$ accounting for the effect of the TF metals in series with one with capacitance per unit area $\varepsilon_0\varepssol/L$ accounting for the effect of the dielectric slab (see also Ref.~\citenum{scalfi_semiclassical_2020}). For $\ltf=0$, Eq.~\ref{eq:totalcharge} reduces to the result derived for an ideal metal in Ref.~\citenum{girotto_simulations_2017}.

The effective length $\leff$ defined in Eq.~\ref{eq:Leff} can thus be understood as the width of an equivalent dielectric slab, and $E_{\rm eff}=-\Delta\Psi/\leff$ is the corresponding uniform electric field in the slab induced by the applied voltage. This effective length is also consistent with the one introduced to describe the lateral decay of the charge induced by an ion in a single TF electrode, proposed recently in Ref.~\citenum{nair_induced_2025} as an extension of earlier work by Vorotyntsev and Kornyshev~\cite{vorotyntsev_electrostatic_1980}, even though here the distance is the one between the two electrodes (hence the factor of two for their contribution) and not an ion-surface distance. Finally, Eq.~\ref{eq:totalcharge} can be rearranged as
\begin{align}
    M_{\rm elec} + M_{\rm ions} + M_{\rm solv} &= 0 \; ,
\end{align}
where we have introduced the dipole of the electrode charge distribution $M_{\rm elec}=Q\leff$ and the solvent dipole $M_{\rm solv}=\varepsilon_0\varepssol L_xL_yL E$, with $E=-\Delta\Psi/L$. This underlines the various contributions to the screening of the electric field throughout the system, as discussed in the case of $\ltf=0$ for explicit electrodes and solvent (without ions)~\cite{pireddu_frequency-dependent_2023}.

\subsubsection{Effective many-body potential between ions}
\label{sec:theory:BO:Veff}

As explained in Eq.~\ref{eq:Fdef}, we evaluate the thermodynamic potential $\Omega[n_{\rm l}^{\rm BO},n_{\rm r}^{\rm BO},\rho_{\rm ions}]$ for the charge density inside the electrodes satisfying the above Euler-Lagrange equations. As a result, it only depends on the ion distribution and, parametrically, on the electron chemical potentials. Using Eqs.~\ref{eq:Omega} and~\ref{eq:Tapprox}, as well as Eq.~\ref{eq:mumuTPsi}, Eq.~\ref{eq:EL2} and the electroneutrality condition, we obtain:
\begin{align}
	V^{\rm eff}[\{\bfr_i\}|\mu_{\rm l},\mu_{\rm r}] &\equiv\Omega[n_{\rm l}^{\rm BO},n_{\rm r}^{\rm BO}|\rho_{\rm ions},\mu_{\rm l},\mu_{\rm r}] \nonumber \\
	& \hspace{-2cm} = \int_{\rm l,r} \left[ \tau(n_{\rm l,r}^0) - \mu_{\rm l,r} n_{\rm l,r}^0 \right] \dbfr
	+ \frac{1}{2}\int_{\rm s}\rho_{\rm ions}\phi \dbfr
        \nonumber \\ &  \hspace{-1cm} +\int_{\rm l,r} \left[ \mu_{\rm l,r}^T - \mu_{\rm l,r}  + \frac{1}{2}\left( \alpha_{\rm l,r}\delta n_{\rm l,r}- e \phi \right)  
	\right] \delta n_{\rm l,r} \dbfr
	\nonumber \\
	& \hspace{-2cm} = \Omega_0 
	+ \frac{1}{2}\int_{\rm s}\rho_{\rm ions}\phi \dbfr + \int_{\rm l,r} \frac{e}{2}(\Psi_{\rm l,r}-\lambda) \delta n_{\rm l,r} \dbfr
    \nonumber \\
	& \hspace{-2cm} = \Omega_0 
	+ \frac{1}{2}\int_{\rm s}\rho_{\rm ions}\phi \dbfr- \frac{1}{2} \left[ (\Psi_{r}-\lambda) Q_{\rm r} + (\Psi_{l}-\lambda) Q_{\rm l} \right] \, .
\label{eq:Omega2}
\end{align}
where $\Omega_0=\displaystyle\int_{\rm l,r} \left[ \tau(n_{\rm l,r}^0) - \mu_{\rm l,r} n_{\rm l,r}^0 \right]\dbfr$ is the electronic grand-potential in the absence of ions and voltage, which does not depend on the ionic configuration and therefore doesn't contribute to the force on the ions. We remind the reader that Eq.~\ref{eq:Omega2} corresponds to an expansion of the kinetic energy to second order in $\delta n_{\rm l,r}$, whereas the Coulomb contribution is exactly quadratic in these quantity. The last term can be rewritten, when the electrolyte is neutral ($\int_{\rm s}\rho_{\rm ions}\dbfr=\sum_i q_i=0$), by introducing the charge of the right electrode $Q=Q_{\rm r}=-Q_{\rm l}$, whose expression is provided in Eq.~\ref{eq:totalcharge}, the voltage $\Delta\Psi$, and $\phi=\phi_{\Delta\Psi}+\phi_{\rm ions}$, as:
\begin{align}
    V^{\rm eff}[\{\bfr_i\}|\Delta\Psi] & = \Omega_0 + \frac{1}{2}\int_{\rm s}\rho_{\rm ions}(\bfr)\phi(\bfr)\dbfr - \frac{1}{2}Q\Delta\Psi 
	\nonumber \\
    & \hspace{-1.5cm} = \Omega_0 -\frac{1}{2}C_{0}\Delta\Psi^2 + \frac{1}{2}\sum_i q_i\phi_{\rm ions}(\bfr_i) + \sum_i q_i\phi_{\Delta\Psi}(\bfr_i) \nonumber \\
	& \hspace{-1.5cm} = \Omega_0 -\frac{1}{2}C_{0}\Delta\Psi^2 + U_{\rm Coul}^{\rm ions}  + \frac{\Delta\Psi}{\leff}M_{\rm ions} \; ,
	\label{eq:Veff}
\end{align}
with 
\begin{align}
    \label{eq:UCoulIons}
    U_{\rm Coul}^{\rm ions}[\{\bfr_i\}]\equiv \frac{1}{2}\sum_i q_i\phi_{\rm ions}(\bfr_i)
\end{align}
where, as mentioned previously, the diverging self-energy $q_i^2/4\pi\varepsilon_0\varepssol|\bfr-\bfr_i|$ is implicitly removed when evaluating $\phi_{\rm ions}(\bfr_i)$.

Note that a constant shift of all electrostatic potentials does not change the r.h.s. of Eq.~\ref{eq:Veff} since it would cancel in $\Delta\Psi$ and would only add a contribution proportional to $\sum_iq_i=0$ due to the electroneutrality of the electrolyte. In fact, this remark also holds more generally in Eq.~\ref{eq:Omega2} due to the global electroneutrality ($Q_{\rm l}+Q_{\rm r}+\sum_iq_i=0$). In addition, the first two terms in Eq.~\ref{eq:Veff} do not depend on the ionic positions and therefore do not contribute to the forces acting on them (see Eq.~\ref{eq:Feff}). The effect of ion-ion interactions in the presence of the implicit solvent and Thomas-Fermi electrodes will be discussed in more detail in Section~\ref{sec:theory:sim:Veff}. Finally, the force acting on ion $i$ due to the applied voltage (negative gradient with respect to $\bfr_i$ of the last term) is simply $q_iE_{eff}\bfe_z$ with the effective field $E_{\rm eff}=-\Delta\Psi/\leff$ introduced in Section~\ref{sec:theory:BO:Charge}.

\subsubsection{Partition function, average charge and capacitance}
\label{sec:theory:BO:StatMech}

Brownian dynamics simulations of ions using the effective many-body potential Eq.~\ref{eq:Veff} sample the statistical ensemble corresponding to fixed number of ions $N_{\rm ions}$, volume $V$, temperature $T$ and voltage $\Delta\Psi$. The corresponding partition function is
\begin{align}
    \mathcal{Z}[\Delta\Psi] = \int {\rm d}\bfr^{N_{\rm ions}} \, e^{-\beta 	V^{\rm eff}[\{\bfr_i\}|\Delta\Psi] }
    \label{eq:defZ}
\end{align}
where $\beta=1/k_BT$ with $k_B$ Boltzmann's constant, and the corresponding thermodynamic potential is the free energy $\mathcal{F}[\Delta\Psi]=-k_BT\ln\mathcal{Z}[\Delta\Psi]$ (see Ref.~\citenum{scalfi_charge_2020}).
The average charge, for a given voltage, follows straightforwardly from Eq.~\ref{eq:totalcharge}:
\begin{align}
    \left\langle Q \right\rangle &= 
    \frac{1}{\mathcal{Z}}\int {\rm d}\bfr^{N_{\rm ions}} \, Q[\{\bfr_i\}|\Delta\Psi]  \,e^{-\beta 	V^{\rm eff}[\{\bfr_i\}|\Delta\Psi] }
    \nonumber \\
    &= C_{0}\Delta\Psi - 	\frac{\left\langle M_{\rm ions} \right\rangle}{\leff}
    \;, 
    \label{eq:avgQ}
\end{align}
where $\left\langle M_{\rm ions} \right\rangle$ is the ensemble-averaged dipole of the ionic distribution. Furthermore, the charge for fixed configuration of the ions and voltage, $Q[\{\bfr_i\}|\Delta\Psi]$ is equal to $-\partial V^{\rm eff}[\{\bfr_i\}|\Delta\Psi]/\partial\Delta\Psi$ (see Eqs.~\ref{eq:Veff} and~\ref{eq:totalcharge}), so that:
\begin{align}
    \left\langle Q \right\rangle
    &= \frac{1}{\mathcal{Z}}\left( -\frac{1}{\beta} \frac{\partial}{\partial\Delta\Psi}\int {\rm d}\bfr^{N_{\rm ions}} \, \,e^{-\beta 	V^{\rm eff}[\{\bfr_i\}|\Delta\Psi] } \right)
    \nonumber\\
    &= \frac{1}{\beta} \frac{\partial \ln\mathcal{Z}}{\partial\Delta\Psi} = - \frac{\partial \mathcal{F}}{\partial\Delta\Psi} \; ,
    \label{eq:avgQ2}
\end{align}
similarly to what was found for simulations with explicit electrodes in the constant-potential ensemble~\cite{scalfi_charge_2020}. This shows that the free energy change during the charge of the capacitor from zero voltage to finite voltage $\Delta\Psi_{\rm max}$ is 
\begin{align}
    \Delta\mathcal{F} &= - \int_0^{\Delta\Psi_{\rm max}} \left\langle Q \right\rangle_{\Delta\Psi} {\rm d}\Delta\Psi \; ,
    \label{eq:DeltaF}
\end{align}
\textit{i.e.} the reversible electrical work exchanged with the charge reservoir. The subscript highlights here the fact that the ensemble average of the charge is made at fixed voltage $\Delta\Psi$.

The differential capacitance is the derivative of the average electrode charge with respect to voltage:
\begin{align}
    C_{\rm diff} = \frac{\partial \left\langle Q \right\rangle}{\partial\Delta\Psi} &= \beta  \left[ \left\langle Q^2 \right\rangle - \left\langle Q \right\rangle^2 \right]
    \nonumber \\ 
    & = C_{0} + \frac{\beta }{\leff^2}
    \left[ \left\langle M_{\rm ions}^2 \right\rangle - \left\langle M_{\rm ions} \right\rangle^2 \right] \;.
    \label{eq:Cdiff}
\end{align}
For a purely capacitive system, this quantity is independent of voltage and equal to the integral capacitance $C_{\rm int}=\left\langle Q \right\rangle/\Delta\Psi$. However this is not necessarily the case in general. Eq.~\ref{eq:Cdiff} is a fluctuation-dissipation relation relating the response of the average electrode charge to a change in voltage to the equilibrium fluctuations of the electrode charge at fixed voltage. These fluctuations arise from two contributions: the thermal fluctuations of the ions, as obvious from the second term in the last line of Eq.~\ref{eq:Cdiff}, and the thermal fluctuations of the implicit electrons and the solvent, which are suppressed in the Born-Oppenheimer description and embedded in $C_0$, via the Thomas-Fermi screening length $\ltf$ and the relative permittivity $\varepssol$, respectively (see Eq.~\ref{eq:capaempty}). This result is analogous to the one obtained  in Ref.~\citenum{scalfi_charge_2020} with explicit electrode atoms and electrolyte (including the solvent). In the latter case, the explicit expression for the contribution of the electrolyte-free capacitor has the same physical origin as $C_0$ but is not as simple as Eq.~\ref{eq:capaempty}, as it depends on the microscopic details of the electrodes.

\subsection{Brownian Dynamics simulations}
\label{sec:theory:sim}

The ions are described as Brownian particles with a diffusion coefficient $D_i$. Their positions $\bfr_i$ evolve according to the overdamped Langevin equation: 
\begin{equation}
    \dot{\bfr}_i =  \beta D_i {\bf F}_i + \sqrt{2 D_i}\bm{\xi}_i
    \label{eq:BD}
\end{equation}
where $\bm{\xi}_i$ is a Gaussian white noise, and the force ${\bf F}_i$ acting on the ions includes the interactions between them and with the walls. They include short-range forces, described in Section~\ref{sec:theory:sim:sr}, as well as the one deriving from the effective many-body potential (see Eq.~\ref{eq:Feff}), described further in Section~\ref{sec:theory:sim:Veff}.

\subsubsection{Short-range interactions}
\label{sec:theory:sim:sr}

Short-range repulsion interactions between ions $i$ and $j$ are described by Weeks-Chandler-Andersen (WCA) potentials:
\begin{equation}
\label{eq:WCA}
v_{ij}^{\mathrm{WCA}}(r) =
    \left\lbrace
    \begin{matrix}
    v_{ij}^\mathrm{LJ}(r) \, - \, v_{ij}^\mathrm{LJ}(r^*) \ &, \ r \leq r^*, \\
    0 \ &, \ r > r^*,
    \end{matrix}
    \right.
\end{equation}
with the Lennard--Jones (LJ) potential
\begin{equation}
\label{eq:LJ}
v_{ij}^\mathrm{LJ}(r) =
4\epsilon_{ij} \left[ \left(\frac{\sigma_{ij}}{r}\right)^{12} -  \left(\frac{\sigma_{ij}}{r}\right)^6 \right],
\end{equation}
and $r^*=2^{1/6}\sigma_{ij}$ the position of the minimum of $v_{ij}^\mathrm{LJ}$. The LJ energy and diameter $\epsilon_{ij}$ and $\sigma_{ij}$ are computed from the corresponding parameters for ions $i$ and $j$ using the Lorentz--Berthelot mixing rules. 

Two descriptions are considered for the short-range interactions between ions and the electrodes. For the comparison with explicit electrodes (see Section~\ref{sec:theory:sim:BD}), we use the same short-range interactions, \textit{i.e.} pairwise WCA interactions with explicit atoms, so that only the many-body effective potential describing Coulomb interactions and the effect of the kinetic energy of the electrons differ from the explicit case. For the other cases, the short-range interactions with the wall are also treated implicitly via a potential of the form
\begin{equation}
    U(z) \ = \ V_\mathrm{w}( z + L/2) \, + \, V_\mathrm{w}( z - L/2),
    \label{eq:Walls}
\end{equation}
which depends only on $z$, where $V_\mathrm{w}$ is the so-called Steele potential, obtained by integrating LJ interactions over atomic planes. The resulting potential includes short-range repulsion and an attractive well leading to adsorption~\cite{steele_physical_interaction_of_gases_1973,steele_interaction_of_rare_gas_1978,magda_molecular_1985,magda_erratum_1986}.


\begin{align}
V_\mathrm{w}^\mathrm{Steele}(z) &=
2 \, \pi \,  \rho_\mathrm{surf}\,\epsilon_\mathrm{w} \sigma_\mathrm{w} \,\Delta\, \displaystyle
\left[
\frac{2}{5} \left( \frac{\sigma_\mathrm{w}}{z}\right)^{10} \,
\right. \nonumber \\ 
 & \left. \hspace{1.5cm} \displaystyle
 - \left( \frac{\sigma_\mathrm{w}}{z}\right)^{4}
 - \frac{\sigma_{w}^4}{3 \Delta (z + 0.61 \Delta)^3}\right]
\ ,
\label{eq:Walls:Steele}  
\end{align}
with  $\rho_\mathrm{surf}$ the surface number density of electrode atoms, $\Delta$ the distance between atomic planes within the electrode, and where $\epsilon_{\rm w}$ and $\sigma_{\rm w}$ tune the strength and range of the ion-wall interaction. In order to model purely repulsive walls, we use the same truncation and shifting as for the short-range ion-ion interactions, \textit{i.e.}
\begin{equation}
    V_{w}(z)  \ = \ 
    \left\lbrace
    \begin{matrix}
    V_\mathrm{w}^\mathrm{Steele}(z) \, - \, V_\mathrm{w}^\mathrm{Steele}(z^*) \ &, \ z \leq z^*, \\
    0 \ &, \ z > z^*,
    \end{matrix}
    \right.
\label{eq:Walls:Vrep} 
\end{equation}
with $z^*\approx 0.986\sigma_\mathrm{w}$. More details on interactions are provided in Section~\ref{sec:theory:sim:BD}.

\subsubsection{Effective many-body potential}
\label{sec:theory:sim:Veff}

The force arising from electrostatic interactions and the kinetic energy of the electrons is computed as the negative gradient with respect to $\bfr_i$ Eq.~\ref{eq:Feff} of the effective many-body potential. As already mentioned in Section~\ref{sec:theory:BO:Veff}, the first two terms in Eq.~\ref{eq:Veff} do not contribute to the force acting on ion $i$. The last term provides the force due to the applied voltage and is simply $q_iE_{eff}\bfe_z$ with the effective field $E_{\rm eff}=-\Delta\Psi/\leff$ introduced in Section~\ref{sec:theory:BO:Charge}. The force due to the effect of ion-ion interactions in the presence of the implicit solvent and Thomas-Fermi electrodes (third term in Eq.~\ref{eq:Veff}) is the negative gradient of $U_{\rm Coul}^{\rm ions}[\{\bfr_i\}]$ defined in Eq.~\ref{eq:UCoulIons} where the diverging self-energy $q_i^2/4\pi\varepsilon_0\varepssol|\bfr-\bfr_i|$ is implicitly removed when evaluating $\phi_{\rm ions}(\bfr_i)$.

The electrostatic energy $U_{\rm Coul}^{\rm ions}[\{\bfr_i\}]$ can be split into two contributions: one due to direct Coulomb interactions between the ions without electrodes (corresponding to the limit of infinite interelectrode distance $L\to\infty$), which can be efficiently calculated under 2D periodic boundary conditions using Ewald summation techniques~\cite{mazars_yukawa_2007a, mazars_yukawa_2007b}, and another due to the polarization of the electron charge distribution inside the metal. Specifically, we write
\begin{align}
    \label{eq:UCoulIonSepar}
    U_{\rm Coul}^{\rm ions} = U_{\rm Coul}^{\infty} + \delta U_{\rm Coul}^{\rm ions}
\end{align}
with (see also Appendix~\ref{sec:appendix:phi}):
\begin{align}
    \label{eq:UCoulIonsLinf}
    U_{\rm Coul}^{\infty}[\{\bfr_i\}] &\equiv \lim_{L\to\infty} \frac{1}{2}\sum_i q_i\phi_{\rm ions}(\bfr_i) \nonumber \\
    &= \frac{1}{2}\sum_i \sum_{j \neq i} \sideset{}{'}\sum_{\bf n}\frac{q_iq_j}{4\pi\varepsilon_0\varepssol|\bfr_j-\bfr_i+{\bf n}|} \; ,
\end{align}
where ${\bf n}=m_xL_x\bfex + m_yL_y\bfey$ accounts for the sum over periodic images and the prime now explicitly indicates the removal of the term $j=i$ when ${\bf n}={\bf 0}$.
Using Eq.~\ref{eq:phi:phiions} for the potential $\phi_{\rm ions}(\bfr)$ with Eq.~\ref{eq:phi:phiionsmodess} for the Green's function in the electrolyte, as well as the electroneutrality of the electrolyte ($\sum_i q_i=0$), we obtain the following expression for the difference:
\begin{widetext}
\begin{align}
    \label{eq:deltaUCoulIons}
    \delta U_{\rm Coul}^{\rm ions} &\equiv U_{\rm Coul}^{\rm ions} - U_{\rm Coul}^{\infty}
    =  \displaystyle \frac{1}{2L_xL_y} \left\{ \sum_i \sum_j \sum_{\bfk\neq{\bf 0}} q_i q_j \delta g_{\mathbf{k}}^s(z_i,z_j)\cos\left[\mathbf{k}\cdot(\bfr_j-\bfr_i)\right] + \frac{M_{\rm ions}^2}{\varepsilon_0\varepssol\leff}\right\}
\end{align}
with (see Eqs.~\ref{eq:phi:phiionsmodess} and~\ref{eq:phi:phiionsmodesLinf})
\begin{align}
	\label{eq:deltags}
	 \delta g_{\mathbf{k}}^s(z_i,z_j)=& \displaystyle  \frac{\left[e^{-k(z_j+z_i+L)}+e^{k(z_j+z_i-L)}\right]\left( \varepssol^2  k^2 -\chi_{TF}^2\right)+\left[e^{k(z_i-z_j-2L)}+e^{k(z_j-z_i-2L)}\right]\left( \varepssol   k-\chi_{TF}\right)^2 }{2\varepsilon_0\varepssol k \left[(1-e^{-2kL})(\varepssol ^2k^2+\chi_{TF}^2)+2(1+e^{-2kL})\varepssol k\chi_{TF}\right]} 
\end{align}
\end{widetext}
where $\chi_{TF}$ is defined in Eq.~\ref{eq:defchiTF} and the last term in Eq.~\ref{eq:deltaUCoulIons} corresponds to $\bf k=0$. Since $z_i$ and $z_j$ are in the interval $]-\frac{L}{2},\frac{L}{2}[$, the sum over modes $\bf k$ converges exponentially fast, regardless of the positions of the two ions. Note that there is no diverging self-energy term to be subtracted in Eq.~\ref{eq:deltaUCoulIons}.

\subsubsection{Simulation details}
\label{sec:theory:sim:BD}

The coupled equations Eq.~\ref{eq:BD} are solved numerically using the MetalWalls simulation package~\cite{marin-lafleche_metalwalls_2020,coretti_metalwalls_2022}, in which we implemented the Euler–Maruyama integrator, a rescaling of the 2D Ewald summation by the permittivity $\varepssol$ to compute $U_{\rm Coul}^{\rm ions}$ (see Eq.~\ref{eq:UCoulIons}) and the new term $\delta U_{\rm Coul}^{\rm ions}$ (see Eq.~\ref{eq:deltaUCoulIons}) to account for the effect of the implicit Thomas-Fermi electrodes, as well as the corresponding forces. 

The parameters for the short-range interactions (see Section~\ref{sec:theory:sim:sr}) are identical for cations, anions and wall atoms, namely $\sigma_{ij}=\sigma_\mathrm{w}=5$~\AA\ and $\epsilon_{ij}=\epsilon_\mathrm{w}=2.477$~kJ/mol. The cut-off for ion-ion interactions is thus $r^* =5.61$~\AA, while that for the Steele potential is $z^*=4.92$~\AA. For the definition of the latter, the graphite structure corresponds to $\rho_{\rm surf}=0.38$~\AA$^{-2}$ and $\Delta=3.354$~\AA. The repulsive walls (explicit atom sites or zero of the Steele potential) are placed at the same position as the dielectric interface, \textit{i.e.} $z=\pm L/2$.

For the Coulomb interactions, we use a real-space cut-off of $r_c=15.9$~\AA\ in the Ewald summation and choose the number of wave vectors to compute 
$U_{\rm Coul}^{\infty}$ and $\delta U_{\rm Coul}^{\rm ions}$ to achieve a prescribed tolerance $\delta U_{\rm tol}=3\times10^{-5}$~eV. For $\delta U_{\rm Coul}^{\rm ions}$, which is computed only in reciprocal space, this leads to $k_{max}>\log\left[\frac{q_{max}^2}{8\pi d \varepsilon_0\varepssol\delta U_{\rm tol}}\right]/2d$, with $d$ the distance of closest approach of ions from the interface. For isolated ions (Section~\ref{sec:results:isolated}) and ion pairs (Section~\ref{sec:results:pair}) we can choose this according to the range of considered ion-distance surfaces. For the Brownian dynamics simulations of Sections~\ref{sec:results:explicit} and~\ref{sec:results:ltf}, we choose $d=\sigma/2$, with $\sigma$ the common diameter describing short-range interactions between ions and with the wall. Except for the results in vacuum for validation on isolated ions and ion pairs, the simulated systems correspond to a solvent with the relative permittivity of water, $\varepssol=78$.


For BD simulations, we consider ions with a diffusion coefficient $D_\pm=1.12\times10^{-9}$~m$^2$s$^{-1}$ and a temperature $T=298$~K. The overdamped Langevin equations~\ref{eq:BD} are integrated with a time step of 5~fs. The systems are first equilibrated for 100~ns, before production runs of 400~ns, with electrode charges (computed from Eq.~\ref{eq:totalcharge}) sampled every 0.25~ps and ion positions every 0.5~ps. The reported uncertainty on ionic concentration profiles is the standard error over 10 consecutive blocks of the trajectory. The differential capacitance is computed from the variance of the charge distribution (first line in Eq.~\ref{eq:Cdiff}). The corresponding uncertainty is obtained as in Ref.~\citenum{scalfi_charge_2020} following Zwanzig and Ailawadi~\cite{zwanzig_statiscal_1969}: The standard error estimated as $\langle\delta Q^2\rangle\times\sqrt{4\tau_{\rm c}/\tau_{\rm s}}$ where $\tau_{\rm s}$ is the sampling time and $\tau_{\rm c}$ is the correlation time computed as $\tau_{\rm c}=\int_0^{\infty}\langle\delta Q(t)\delta Q(0)\rangle^2/\langle\delta Q^2\rangle^2 \diff t$ . The uncertainty over the integral capacity is computed via the same method with $\langle Q\rangle\times\sqrt{4\tau'_{\rm c}/\tau_{\rm s}}$ with $\tau'_{\rm c}=\int_0^\infty\langle\delta Q(t)\delta Q(0)\rangle/\langle\delta Q^2\rangle\diff t$. We refer the reader to Ref.~\citenum{pireddu_impedance_2024} for more details on the computation of these correlation times.

\section{Results}
\label{sec:results}

We now demonstrate the relevance of the proposed implicit electrode description, by first validating its predictions for the force on the ions as a function of the ion-surface distance in simple geometries: we compare our results with (semi-)analytical for isolated ions near electrodes for various $\ltf$ in Section~\ref{sec:results:isolated}) and with numerical results under periodic boundary conditions for ion pairs next to perfect metals ($\ltf=0$) in Section~\ref{sec:results:pair}. We then turn in Section~\ref{sec:results:explicit} to the equilibrium ionic density profiles and compare our results with Brownian Dynamics simulations with an explicit electrode model for perfect metals proposed in Ref.~\citenum{cats_capacitance_2022}, before using our model that also applies to finite $\ltf$ to discuss the effect of the Thomas-Fermi screening length on the ionic density profiles in Section~\ref{sec:results:ltf}. Section~\ref{sec:results:capa} finally discusses the capacitance as a function of $\ltf$ and compares, when possible, several levels of description.

\subsubsection{Isolated ions}
\label{sec:results:isolated}

As a first test of the present method, we compare our numerical predictions for a periodic system with two electrodes and several ions to the semi-analytical results from the literature for a single charge in a dielectric next to a Thomas-Fermi metal. We use the result for this non-periodic case with a single solid/liquid interface in a superposition approximation to compare with our case with two electrodes and several ions, without considering the effect of periodic boundary conditions. Specifically, we compare $U_{\rm Coul}^{\rm ions}[\left\{\bfr_i\right\}]$ for a simple geometry with the reference :
\begin{align}
\label{eq:UCoulRef:tot}
    U_{\rm Coul}^{\rm ref}[\left\{\bfr_i\right\}]& \equiv \frac{1}{2} \sum_i q_i \phi_{i}^{\rm ref}(\bfr_i) = \sum_i U_{i}^{\rm ref}(\bfr_i) 
\end{align}
with 
\begin{align}
\label{eq:UCoulRef:i}
    U_{i}^{\rm ref} (\bfr_i) & = \displaystyle \frac{q_i^2}{8\pi\varepsilon_0\varepssol} \displaystyle\int_0^\infty \mathrm{d}k\frac{\varepssol k-\sqrt{k^2+k_{TF}^2}}{\varepssol k+\sqrt{k^2+k_{TF}^2}}e^{-2kd_i}
\end{align}
where $d_i$ is the distance of ion $i$ from the nearest electrode. Eq.~\ref{eq:UCoulRef:i} was first derived by Kornyshev \textit{et al.} (see Eqs.~6 and following of Ref.~\citenum{kornyshev_image_1977}) and used as reference in several subsequent works (see in \emph{e.g.} Eq.~S19 of Ref.~\citenum{schlaich_electronic_2022} or more recently Eq.~69 of Ref.~\citenum{hedley_what_2025}). This superposition approximation assumes in particular that $d_i\ll L$ for all ions and that the lateral dimensions are sufficiently large $d_i\ll L_x,L_y$. Eq.~\ref{eq:UCoulRef:i} reduces to the well-known limits:
\begin{align}
\label{eq:UCoulRef:i:ltf0}
    \lim_{\ltf\to0}U_{i}^{\rm ref} (\bfr_i) & = -\displaystyle \frac{q_i^2}{16\pi\varepsilon_0\varepssol d_i}
\end{align}
and
\begin{align}
\label{eq:UCoulRef:i:ltfinfinity}
    \lim_{\ltf\to\infty}U_{i}^{\rm ref} (\bfr_i) & = \displaystyle \left(\frac{\varepssol-1}{\varepssol+1}\right)\frac{q_i^2}{16\pi\varepsilon_0\varepssol d_i}
\end{align}
for perfect conductors and insulating walls, respectively.

\begin{figure}[ht!]
    \begin{center}
        \includegraphics[width=0.45\textwidth]{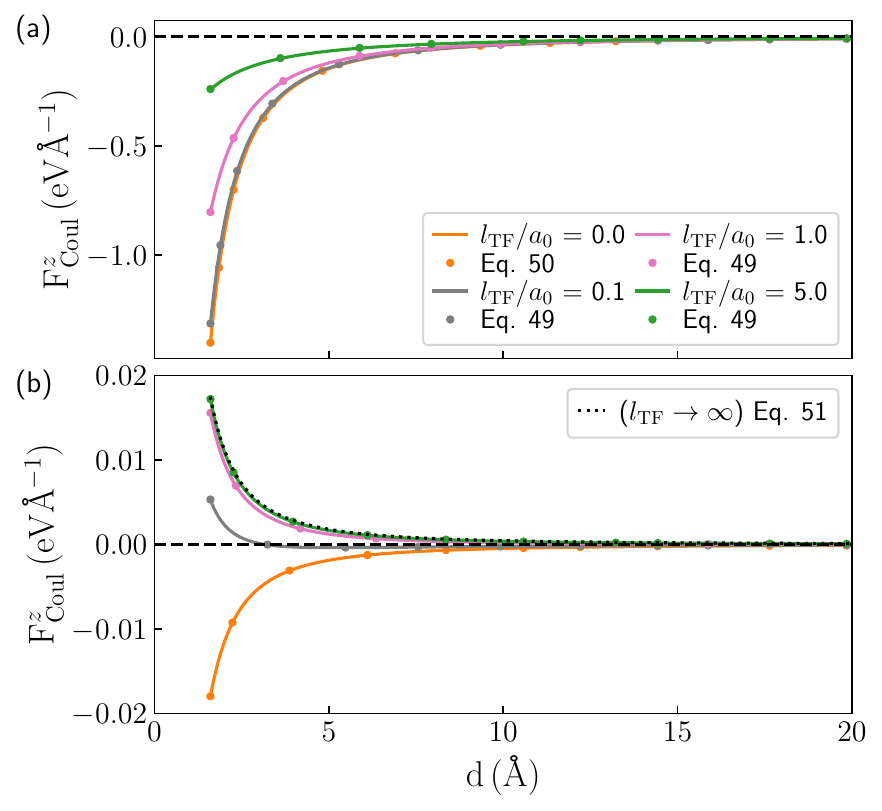} 
    \end{center}
    \caption{
    Electrostatic force in the direction perpendicular to the implicit electrodes, as a function of the distance $d$ from them, for a system consisting of two ``isolated'' ions with charges $+e$ at $(0,0,-\frac{L}{2}+d)$ and $-e$ at $(0,0,+\frac{L}{2}-d)$ in vacuum (a) or in a solvent with relative permittivity $\varepssol=78$ (b), in the absence of voltage. The reported force is that on the positive charge; the force on the negative charge is simply the opposite. The numerical results with the present method (lines) for various Thomas-Fermi screening lengths $\ltf$ (in atomic units, with $a_0\approx0.53$~\AA\, the Bohr radius) indicated by colors are compared with the force corresponding to Eqs.~\ref{eq:UCoulRef:tot} and~\ref{eq:UCoulRef:i}, for independent ions next to an infinite interface between a dielectric medium and a Thomas-Fermi metal (symbols). The dotted line in panel b indicates the insulating limit $\ltf\to\infty$ (see Eq.~\ref{eq:UCoulRef:i:ltfinfinity}).
    }
    \label{fig:isolated}
\end{figure}

Figure~\ref{fig:isolated} compares the energy $U_{\rm Coul}^{\rm ions}$ with $U_{\rm Coul}^{\rm ref}$ for a system with an interelectrode distance $L= 159$~\AA\, and lateral dimensions $L_x=L_y=L$, and two ions with opposite charges $q=\pm e$ placed symmetrically at varying distances $d$ from each electrode, \emph{i.e.} positions $x_i=y_i=0$ and $z_i=\pm(\frac{L}{2}-d)$, in vacuum (panel~\ref{fig:isolated}a) or in a solvent with relative permittivity $\varepssol=78$ (panel~\ref{fig:isolated}b). In both cases, the numerical results for all considered TF screening length $\ltf$ ranging from 0 (perfect metal) to $5a_0$, with $a_0\approx0.53$~\AA\, the Bohr radius, are in excellent agreement with the semi-analytical prediction. This simultaneously shows that the box sizes and considered $d$ range are consistent with the superposition approximation and that the proposed scheme provides the expected results. We observe in particular that, while in vacuum the force is always attractive, decays with the ion-surface distance $d$ and with increasing $\ltf$, the dielectric contrast with the solvent results in a cross-over from purely attractive for a perfect metal ($\ltf=0$) to repulsive for large $\ltf$, as expected for insulating walls ($\ltf\to\infty$). For intermediate $\ltf$, the force profile may be non-monotonic, with both repulsive and attractive forces depending on $d$, as observed here for $\ltf=0.1a_0$.

\subsubsection{Ion pair next to a perfect metal}
\label{sec:results:pair}

\begin{figure}[ht!]
    \begin{center}
        \includegraphics[width=0.45\textwidth]{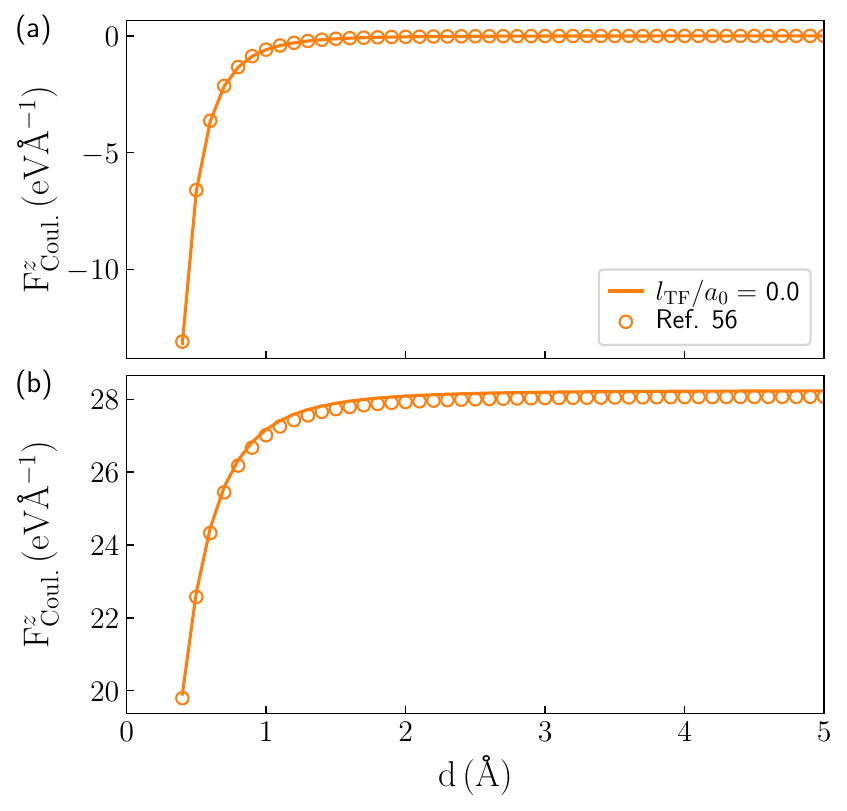} 
    \end{center}
    \caption{
    Electrostatic force in the directions perpendicular (a) and parallel (b) to the implicit electrodes as a function of the distance $d$ from them, for a pair of ions with charges $+e$ at $(0,0,-\frac{L}{2}+d)$ and $-e$ at $(x,0,-\frac{L}{2}+d)$ with $x=0.714$~\AA, in vacuum, for perfect metals ($\ltf=0$). The reported force is that on the positive charge; the force on the negative charge is simply equal for the $z$ component and opposite for the $x$ one. The numerical results with the present method (lines) are compared with that obtained with the method of Ref.~\citenum{telles_efficient_2024} (symbols), which only applies to perfect metals.
    }
    \label{fig:pair}
\end{figure}

We then compare the present method with the existing one for perfect metals ($\ltf=0$) described in Ref.~\citenum{telles_efficient_2024} for a simple system where an ion pair is approached from the same electrode. Specifically, we consider a pair with charges $+e$ at $(0,0,-\frac{L}{2}+d)$ and $-e$ at $(x,0,-\frac{L}{2}+d)$ with $x=0.714$~\AA, in vacuum. The box dimensions are $L_x=67.69$~\AA, $L_y=36.64$~\AA, $L=39.72$~\AA. Fig.~\ref{fig:pair} reports the components of the electrostatic force on the positive charge in the directions perpendicular and parallel to the electrodes as a function of the distance $d$ from them (the force on the negative charge is simply equal for the $z$ component and opposite for the $x$ one). The results for both components as a function of distance from the interface are in excellent agreement with the ones obtained by the method of Ref.~\citenum{telles_efficient_2024}. This proves that the present approach, which is implemented as a correction to the case of ions in the absence of walls under 2D periodic boundary conditions (see Eq.~\ref{eq:UCoulIonSepar}) correctly describes the effect of the interface in the case of a perfect metal ($\ltf=0$).

\subsubsection{Comparison with explicit electrodes and implicit solvent}
\label{sec:results:explicit}

After the above validation for specific configurations of the ions by comparison with previous analytical and numerical results, we now turn to the study of equilibrium properties of capacitors obtained by Brownian dynamics simulations. We compare the results obtained with the present model describing both the polar solvent and the electrodes implicitly via the Green's functions to that obtained in the case of perfect metals ($\ltf=0$) by Cats \textit{et al.} using explicit electrodes~\citenum{cats_capacitance_2022}. In this work, the authors computed the forces on ions to be used in Brownian dynamics (in their case underdamped Langevin dynamics) by adapting the method introduced for molecular dynamics simulations with explicit electrode atoms maintained at a constant potential (with a potential difference between the two electrodes) and an explicit electrolyte (ions and solvent molecules). They suggested that for perfect metals, an implicit solvent can be modeled by simultaneously rescaling the ionic charges by a factor $1/\sqrt{\varepssol}$ and the voltage by a factor  $\sqrt{\varepssol}$. While this rescaling in fact amounts to considering that the electrode atoms are also embedded in a medium with permittivity $\varepssol$, this approach resulted in an excellent agreement for the considered systems (monovalent ions in a very polar solvent) and voltages with classical Density Functional Theory calculations.

Here, we consider exactly the same system as in Ref.~\citenum{cats_capacitance_2022}, namely a capacitor consisting of two graphite electrodes separated by $L=39.72$~\AA. Each electrode is modeled by a single atomic layer of graphite with unit cell parameter $a=2.46$~\AA\ and total dimensions $L_x\times L_y=67.69\times36.64$~\AA$^2$, with a total of 960 atoms. We compare the results of overdamped Langevin dynamics simulations (Eq.~\ref{eq:BD}) with forces computed as in Ref.~\citenum{cats_capacitance_2022} or with the present method, which does not rely on explicit electrode atoms to compute the electrostatic forces on the ions. For the short-range interactions, we consider either explicit atomic sites (interacting with the ions via pairwise WCA potentials, see Eq.~\ref{eq:WCA}) to isolate the effect of the electrostatic interactions, or implicit walls (see Eqs.~\ref{eq:Walls}-\ref{eq:Walls:Vrep}), to isolate the effect of the short-range ones.

\begin{figure}[ht!]
    \begin{center}
        \includegraphics[width=0.45\textwidth]{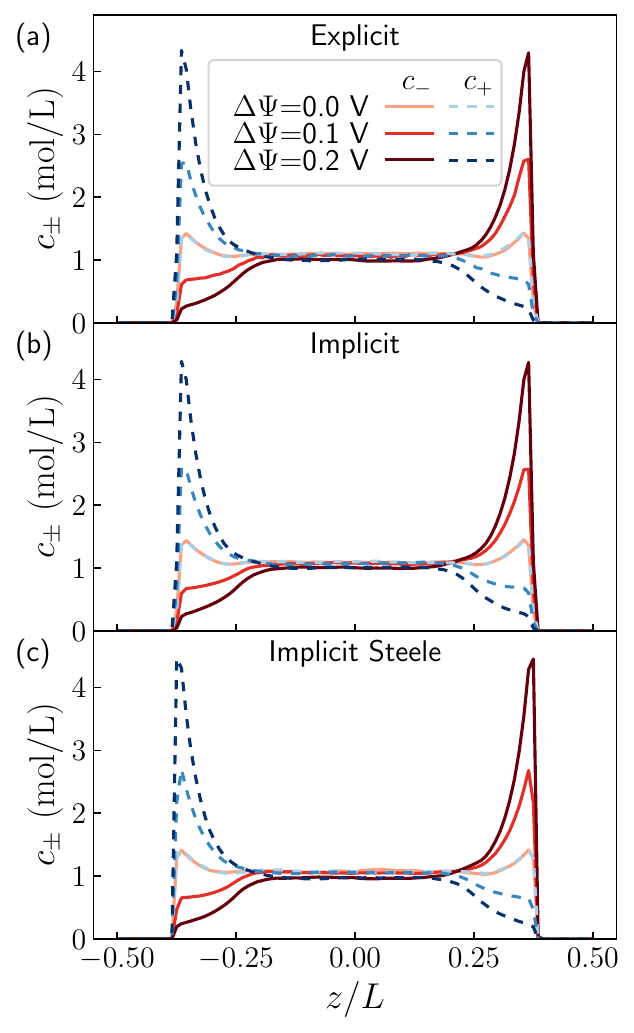} 
    \end{center}
    \caption{
    Equilibrium anion (red solid lines) and cation (blue dashed lines) density profiles from Brownian dynamics simulations of ions in an implicit solvent between perfect metals ($\ltf=0$), with (a) the explicit electrode model from Ref.~\citenum{cats_capacitance_2022} (see text), (b) the present model of implicit electrodes and short-range interactions computed with the same atomic sites as in the previous panel and (c) the present fully implicit model where short-range interactions are computed via the Steele potential (see Eqs.~\ref{eq:Walls}-\ref{eq:Walls:Vrep}). In all panels, the density profiles are shown for $\Delta\Psi=0$, $0.1$ and $0.2$~V from light to dark lines, respectively.
    }
    \label{fig:explicitimplicit}
\end{figure}

The results obtained with the three methods, shown in Fig.~\ref{fig:explicitimplicit}, are in excellent agreement with each other. As expected for an electrolyte consisting of ions with opposite charges and identical short range interactions, the profiles are always symmetric with respect to $z=0$. They are identical in the absence of voltage, and display a small maximum near the electrodes, which results not only from the weak attractive interactions of each ion with the metal (see the orange line in Fig.~\ref{fig:isolated}b for $\ltf=0$) but also from the pressure inside the relatively dense liquid due to short-range repulsion (see also Section~\ref{sec:results:ltf}). As voltage is applied, cations (resp. anions) accumulate at the electrode with the lower (resp. higher) potential and are depleted from the other one, as expected. The accumulation/depletion of ions near the corresponding electrodes increases with applied voltage.

The agreement between the various methods show that the same prediction for the ionic density profiles can be obtained with a significantly simpler model and, accordingly, a reduced computational cost. Indeed, for the considered system with a single layer of electrode atoms in each electrode (as in Ref.~\citenum{cats_capacitance_2022}), with a total of 1920 electrode atoms in the system, the computational cost of electrostatic interactions is reduced by a factor of $\sim 6$ (all comparisons reported here are obtained from simulations with a single CPU on the same computer). For a system with three atomic planes in each electrodes, the speed-up is by a factor of $\sim 60$. For the same systems, the speed-up achieved using an implicit wall for short-range interactions instead of explicit sites is of approximately 25 and 80, respectively. For these systems, the computation of electrostatic interactions represents $\sim 95$\% of the cost per step in the fully implicit model. Finally, while we compared the present implicit electrode model to one with explicit electrodes, in both cases the solvent is treated implicitly. 

The speed-up with respect to a fully atomistic simulation is even more dramatic. As an example, for a system considered in Ref.~\citenum{pireddu_impedance_2024} with a 0.5~M NaCl aqueous solution between model gold electrodes separated by a distance $L\approx5$~nm (with 40 ions, 2160 water molecules and 3240 electrode atoms), the present BD simulations (with only 40 ions and implicit solvent and electrodes) is $\approx 6$ times faster on a single CPU than the MD simulations on a GPU. The present model only approximately describes the effect of the solvent, neglecting in particular subtle effects that may occur in the first molecular layers at the interface, and a detailed comparison remains out of the scope of the present work (see some discussion in Section~\ref{sec:conclusion}). Nevertheless, it offers a computationally efficient approach to investigate the effect of the TF screening length on the properties of capacitors with dimensions much larger than the ones currently within reach of MD simulations. This includes the possibility to consider lower salt concentrations (typically above 0.1~mol/L in MD simulations), by extending the lateral dimensions to ensure a sufficient number of ions in the system, or larger inter-electrode distances (typically limited to 10-20~nm in MD simulations). Importantly, the reduced computational cost per step and the larger time steps used in Brownian dynamics offer the possibility to also investigate the dynamics over much longer times scales (typically limited to a few tens of ns in MD simulations).

\subsubsection{Effect of the Thomas-Fermi screening length}
\label{sec:results:ltf}

We now use the present implicit model to investigate the effect of the Thomas-Fermi screening length on the ionic distributions and the capacitance. We recall that this was not possible with previous methods based on Green's functions, which only applied so far to the perfect metal case~\cite{girotto_simulations_2017,dos_santos_simulations_2017,telles_efficient_2024}. In the following, we consider the same system as in Section~\ref{sec:results:explicit}, except for the variable $\ltf$.

\begin{figure}[ht!]
    \begin{center}
        \includegraphics[width=0.45\textwidth]{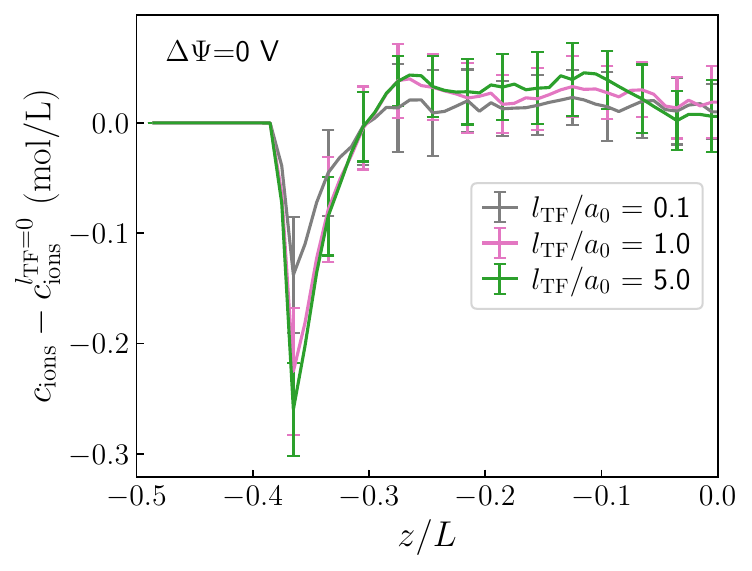} 
    \end{center}
    \caption{
        Difference between the ion concentration profiles ($c_{\rm ions}=c_++c_-$) for various Thomas-Fermi screening lengths $\ltf$ (in units of the Bohr radius $a_0$) and the concentration profiles for $\ltf=0$, in the absence of voltage ($\Delta\Psi=0$). The system is identical to that of Fig.~\ref{fig:explicitimplicit}, except for the change in $\ltf$.
    }
    \label{fig:profilesDPsi0p0V}
\end{figure}

Fig.~\ref{fig:profilesDPsi0p0V} shows the effect of Thomas-Fermi screening lengths $\ltf$ on the ion density profile ($c_{\rm ions}=c_++c_-$), in the absence of voltage ($\Delta\Psi=0$). Specifically, it reports the difference between the profile for a given $\ltf$ with respect to the profile for perfect metals ($\ltf=0$). This reference profile can be seen in Fig.~\ref{fig:explicitimplicit}, where the cation and anion density profiles overlap for $\Delta\Psi=0$~V. The negative values of the difference next to the wall results reflect the cross-over from attractive to repulsive ion-surface interactions, illustrated for isolated ions in panel Fig.~\ref{fig:isolated}b, upon increasing $\ltf$. Note that the decrease in the local ion concentration at the surface remains small (-0.3~mol/L at most for the largest considered $\ltf$, comparable to an insulating wall) compared to the maximum of the concentration ($c_{\rm ions}^{\rm max}\sim2.9\pm0.1$~mol/L, as can be deduced from the ionic density profiles of Fig.~\ref{fig:explicitimplicit}): the balance of interactions, including the short-range repulsion within the electrolyte, still results in density maxima near the walls in that case (see also Fig.~\ref{fig:profilesDPsi0p1V} below). Far from the surface, there is no significant effect of $\ltf$ on the salt concentration, as expected.

\begin{figure}[ht!]
    \begin{center}
        \includegraphics[width=0.5\textwidth]{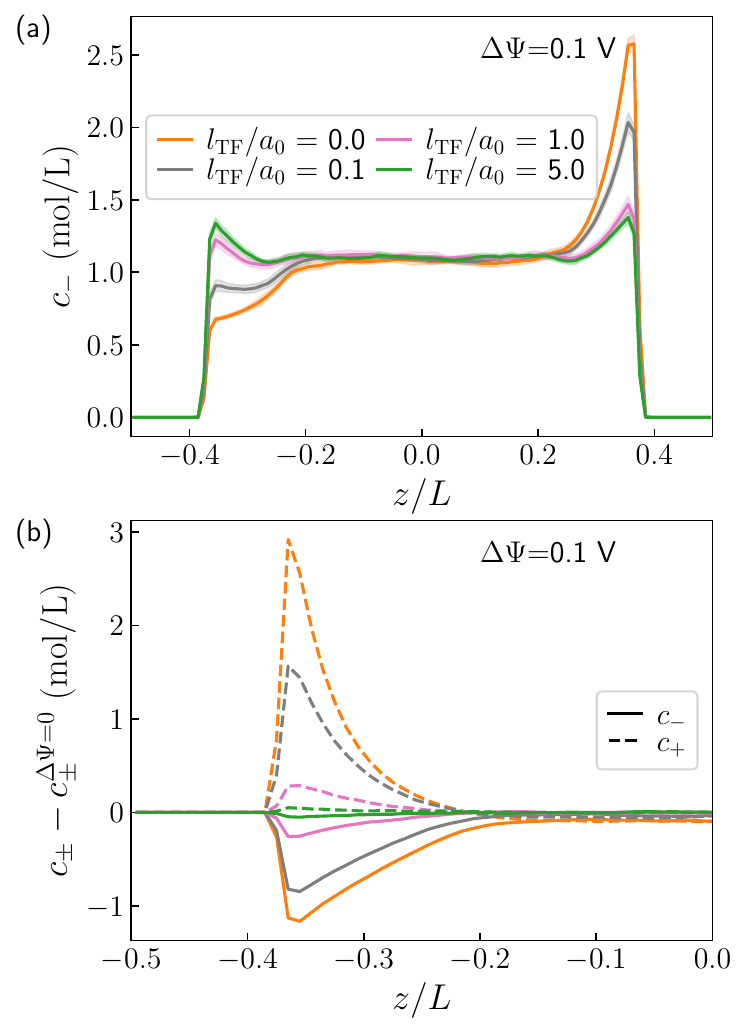} 
    \end{center}
    \caption{
        (a) Anion density profile under an applied voltage $\Delta\Psi=0.1$~V for various Thomas-Fermi screening lengths $\ltf$.
        (b) Difference between the anion (solid lines) and cation (dashed lines) concentration profiles at $\Delta\Psi=0.1$~V with respect to the profiles at $\Delta\Psi=0$~V, for various $\ltf$. The system is identical to that of Fig.~\ref{fig:explicitimplicit}, except for the change in $\ltf$.
    }
    \label{fig:profilesDPsi0p1V}
\end{figure}

Fig.~\ref{fig:profilesDPsi0p1V}a then shows the anion density profiles for a finite voltage $\Delta\Psi=0.1$~V, for various $\ltf$. The cation density profiles, not shown for clarity, are simply symmetric with respect to $z=0$. In all cases, this voltage corresponding to 4 times the thermal potential $k_BT/e\sim25$~mV induces an asymmetry in the anion density profile, with a decrease (resp. increase) with respect to $\Delta\Psi=0$~V next to the left (resp. right) electrode at the lower (resp. larger) potential. This asymmetry is less pronounced as $\ltf$ increases. The effect of voltage is further emphasized in panel~\ref{fig:profilesDPsi0p1V}b, which reports the difference with respect to the cation and anion density profiles for $\Delta\Psi=0$~V, across the left half of the capacitor ($z<0$). The difference highlights the fact that the decrease in cation concentration and increase in anion concentration induced by voltage are symmetric when $\ltf$ is large: in that case the effective electric field inside the capacitor, corresponding to the effective length $\leff$ (see Eq.~\ref{eq:Leff}) is weaker due to screening within the electrodes; this results in small changes in the ionic density profiles, symmetric and decaying approximately exponentially consistently with Debye-H\"uckel (linearized Poisson-Boltzmann) theory, even though one should not expect it to be accurate at such a high concentration. For smaller $\ltf$ values, the lack of screening within the metal results in large and asymmetric depletion/enrichment of the corresponding ions.

\subsubsection{Capacitance}
\label{sec:results:capa}

\begin{figure}[ht!]
    \begin{center}
        \includegraphics[width=0.5\textwidth]{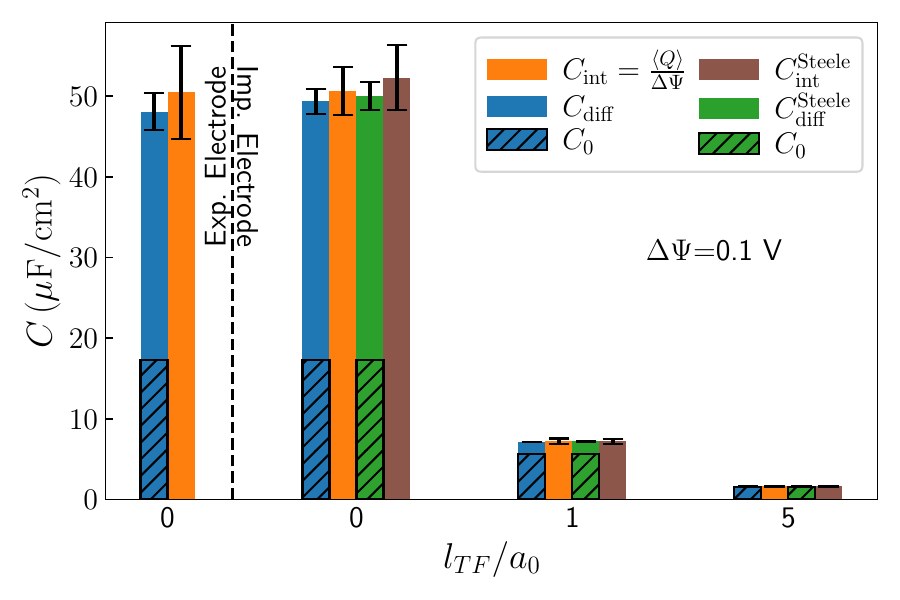} 
    \end{center}
    \caption{
    Integral ($C_{\rm int}$) and differential ($C_{\rm diff}$) capacitance of capacitors measured at $\Delta\Psi=0.1$~V, for capacitors with electrodes characterized by various Thomas-Fermi screening $\ltf$ (in units of the Bohr radius $a_0$), using the implicit electrode model for electrostatic interactions and either explicit or implicit (Steele) description of short-range interactions. For $\ltf=0$, results for the explicit electrode model are also reported. In each case, the contribution $C_0$ of the ion-free capacitor to the differential capacitance (see Eqs.~\ref{eq:capaempty} and~\ref{eq:Cdiff}) is indicated as hatched areas.
    }
    \label{fig:capacitance}
\end{figure}

Finally, Fig.~\ref{fig:capacitance} reports the integral and differential capacitance of capacitors measured at $\Delta\Psi=0.1$~V, for capacitors with electrodes characterized by various Thomas-Fermi screening $\ltf$, using the implicit electrode model for electrostatic interactions and either the explicit or implicit description of short-range interactions. For $\ltf=0$, results for the explicit electrode model are also reported. In all cases, the integral and differential capacitances are in good agreement, as expected for a purely capacitive system. A similarly good agreement is observed between the capacitances predicted with the various descriptions of the same system. This highlights in particular the relevance of the simplest (fully implicit electrodes) introduced in the present work, which provides the same results at a significantly reduced computational cost. 

The capacitance decreases upon increasing $\ltf$, consistently with the cross-over from perfect to insulating walls. Fig.~\ref{fig:capacitance} also indicates the contribution $C_0$ of the ion-free capacitor to the differential capacitance (see Eq.~\ref{eq:Cdiff}). Both $C_0$, corresponding to the fluctuations of the solvent polarization in the absence of ions (see Eq.~\ref{eq:capaempty}) that are taken into account implicitly via the permittivity $\varepssol$, and the remaining contribution corresponding to the ionic contribution to the fluctuations of the dipole of the liquid slab, decrease with increasing $\ltf$. However, the ionic contribution, which dominates for small $\ltf$, becomes negligible for larger values.

\section{Conclusion and perspectives}
\label{sec:conclusion}

We have introduced an efficient description of electrodes, characterized by the Thomas-Fermi screening length $\ltf$ inside the metal, for Brownian dynamics simulations of capacitors. Within a Born-Oppenheimer approximation for the electron charge density inside the electrodes, we derived the effective many-body potential for ions in an implicit solvent between Thomas-Fermi electrodes, taking into account the constraints of applied voltage between them and of global electro-neutrality of the system, as well as the 2D periodic boundary conditions along the electrode surfaces. Following previous work in the case of perfect metals ($\ltf=0$), the problem is solved using Green's functions. The final result is conveniently expressed as a correction to the Ewald summation used to compute ion-ion interactions in standard simulation packages and fully computed in reciprocal space. By considering the statistical ensemble corresponding to fixed number of ions, volume, temperature and voltage, we derived using the effective many-body potential the expressions of the average charge and the fluctuation-dissipation relation for the differential capacitance, highlighting the contributions of ion distribution inside the capacitor and is fluctuations, as well as those from the solvent polarization and of the electron density, whose fluctuations are suppressed within the Born-Oppenheimer description.

We implemented this implicit description of the electrostatic interactions between ions in the presence of the solvent and the electrodes in the MetalWalls simulation package, considering both explicit and implicit descriptions of the short-range interactions between ions and the walls. We demonstrated the relevance of this model by first validating its predictions for the force on the ions as a function of the ion-surface distance in simple geometries, by comparing with results from the literature in limit cases (isolated ions or perfect metals). We then compared the equilibrium ionic density profiles from Brownian Dynamics simulations with those obtained using an explicit electrode model for perfect metals, finding excellent agreement and demonstrating the benefit of using the present implicit model in terms of computational cost.  Finally, we used the present model to discuss the effect of the Thomas-Fermi screening length on the equilibrium ionic density profiles and the capacitance. 

The present model is limited to parallel plate capacitors, although the various steps of Section~\ref{sec:theory} could in principle be extended to undulating surfaces, following Ref.~\citenum{pogharian_electric_2024} where Green's functions were derived for undulating dielectric membranes. But it can be used to investigate the effect of the TF screening length on the properties of capacitors with dimensions much larger than the ones currently within reach of molecular dynamics simulations. This would offer the possibility to consider lower salt concentrations (typically above 0.1~mol/L in MD simulations), by extending the lateral dimensions to ensure a sufficient number of ions in the system, or larger inter-electrode distances (typically limited to 10-20~nm in MD simulations). Importantly, the reduced computational cost per step and the larger time steps used in Brownian dynamics offer the possibility to also investigate the dynamics over much longer times scales (typically limited to a few tens of ns in MD simulations). It would in particular be interesting to derive the fluctuation-dissipation relation for the frequency-dependent admittance (inverse of the impedance), already successfully applied in MD simulations to link the electrochemical response with the dynamics within the liquid slab~\cite{pireddu_frequency-dependent_2023, pireddu_impedance_2024}, in the case of Brownian dynamics, for example following the work of Ref.~\citenum{hoang_ngoc_minh_frequency_2023} for the frequency-dependent conductivity of confined electrolytes. Beyond the study of these effects, the present BD approach can provide reference data to assess the predictions of simpler models such as PNP theory and its extensions, \textit{e.g.} the ones mentioned in the introduction.

In the present work, we only considered systems without redox reactions at the electrode surfaces. In this purely capacitive case, one can assume that the electrolyte is overall neutral (hence that the two electrode charges are opposite) to obtain simple expressions for the electrode charge (Eq.~\ref{eq:totalcharge}) and the many-body effective potential (Eq.~\ref{eq:Veff}). However many steps of the derivation in Section~\ref{sec:theory} do not rely on this assumption. In particular, Eq.~\ref{eq:Omega2} remains valid and one could obtain an expression when the electrolyte is not neutral and its charge is compensated by the net charge of the electrodes. The result could then be used in hybrid simulations where the Brownian dynamics of ions is combined with a Monte Carlo scheme for tentative instantaneous charge transfers between ions and electrodes, as done in Ref.~\citenum{dwelle_constant_2019} for perfect metals. Such an extension would allow to investigate the effect of the Thomas-Fermi length \textit{e.g.} the relaxation of the electric double layer after an electron transfer~\cite{grun_relaxation_2004} or on the correlation between electrodes in electrochemical cells~\cite{huang_correlated_2023}.

Even though the TF model was successfully used to predict the effect of the screening length on a variety of interfacial properties, the accuracy of its predictions for a specific material can be limited, especially for properties that crucially depend on short-range features (expectedly so, since it is the low-wavevector and low-frequency limit of Lindhard theory of screening). Subtle effects, such as the electron spillover at the surface of the metal, which can enhance the capacitance (sometimes described as negative contributions to the inverse capacitance, \textit{e.g.} in jellium models~\cite{schmickler_interphase_1984, leiva_double_1987}), could be approximately captured at the present level by a careful definition of an effective sharp interface between the jellium-like metal and its neighborhood (vacuum or liquid), see \textit{e.g.} Refs.~\citenum{scalfi_charge_2020, nair_induced_2025}. In order to go beyond, one could follow the steps described in Section~\ref{sec:theory} with a better kinetic energy functional. Adding the von Weizs\"acker correction, that involves the gradient of the electronic density~\cite{weizsacker_zur_1935, benguria_thomasfermi_1981}, to the TF term Eq.~\ref{eq:tau} would be a natural step in this direction.

Future possible improvements also pertain to the description of the solvent. The effective interaction between an ion embedded in a dielectric and a Thomas-Fermi metal (see Fig.~\ref{fig:isolated}b) only partly reflects the actual potential of mean force (PMF), which could in principle be computed from all atoms MD simulations with constant-potential TF electrodes~\cite{scalfi_semiclassical_2020}. While a comparison of the present model with an explicit solvent is beyond the scope of the present work, we might anticipate that features such as the molecular layering of the solvent can lead to differences with the predictions of the present implicit model. Some of them can be captured in an effective way by tuning the position of the interface between the metal and the implicit solvent, using the concept of Dielectric Dividing Surface introduced by Netz an co-workers~\cite{schlaich_water_2016} (see also Ref.~\citenum{cox_dielectric_2022}), as illustrated for the frequency-dependent permittivity of nanocapacitors~\cite{pireddu_frequency-dependent_2023,pireddu_impedance_2024}. 

However others, in particular related to the solvation of interfacial ions, might require more elaborate descriptions such as molecular Density Functional Theory~\cite{jeanmairet_molecular_2013, jeanmairet_study_2019, nair_ions_2025}. Such molecular descriptions would however not allow to simply express the many-body effective potential between ions. This could however be achieved by introducing the solvent polarization as an additional field in the free energy functional and treating this polarization at the Born-Oppenheimer level, as the electron charge density. In fact, the present model can be recovered in such a perspective using the simplest functional corresponding to a linear dielectric medium characterized by its permittivity. More elaborate functionals for the solvent (see \emph{e.g.} Refs.~\citenum{berthoumieux_dielectric_2019, becker_dielectric_2025, hedley_what_2025}) could then be introduced naturally at this level of description. In practice, a simpler approach would be to introduce molecular features through a one-body ion-wall PMF, derived or inspired from MD simulations~\cite{loche_effects_2022}. Such terms (\textit{e.g.} with damped oscillations to capture the solvent layering), have already been used in extensions of Poisson-Boltzmann theory~\cite{horinek_specific_2007} or in simulations of electrokinetic phenomena~\cite{joly_liquid_2006} and could straightforwardly be included in the present BD simulations via additional contributions to Eq.~\ref{eq:Walls}.

\section*{Acknowledgments}

This article is dedicated to Christoph Dellago, whose work on developing simulation methods based on advanced Statistical Mechanics has been an inspiration over the years, on the occassion of his $60^{th}$ birthday.
The authors acknowledge discussions with Lydéric Bocquet, Benoit Coasne, David Limmer, Leonardo Coello, Mathieu Salanne and Andrea Grisafi. This project received funding from the European Research Council under the European Union’s Horizon 2020 research and innovation program (grant agreement no. 863473). APdS acknowledges financial support from CNPq under grant 303310/2025-1.

\section*{Author declarations}

\subsection*{Conflict of interest}
There is no conflict of interest to declare.

\subsection*{Author contributions}

\textbf{Paul Desmarchelier:} Conceptualization (equal); Formal analysis (equal); Investigation (lead); Methodology (supporting); Software (lead); Writing/Original Draft Preparation (supporting); Writing – review \& editing (equal). \textbf{Alexandre Dos Santos:} Conceptualization (equal); Formal analysis (equal); Investigation (supporting); Methodology (equal); Writing – review \& editing (equal). \textbf{Yan Levin:} Conceptualization (equal); Formal analysis (equal); Investigation (supporting); Methodology (equal); Writing – review \& editing (equal). \textbf{Benjamin Rotenberg:} Conceptualization (equal); Formal analysis (equal); Funding Acquisition (lead); Investigation (supporting); Methodology (equal); Supervision (lead); Writing/Original Draft Preparation (lead); Writing – review \& editing (equal).

\section*{Data availability}

The original data presented in this study are openly available in Zenodo at \url{https://doi.org/10.5281/zenodo.18607455}.



\appendix

\section{Electrostatic potential}
\label{sec:appendix:phi}

The electrostatic potential at position $\bfr$ due to the ion distribution $\sum_i \rho_i(\bfr)$ where $\rho_i(\bfr)$ given in Eq.~\ref{eq:rhoi} corresponds to ion $i$ with charge $q_i$ at position $\bfr_i$ and its periodic images is:
\begin{widetext}
\begin{eqnarray}
\label{eq:phi:phiions}
\phi_{\rm ions}(\bfr)=
    \begin{cases} 
	\displaystyle \frac{1}{L_xL_y}\sum_{i}^{N_{\rm ions}}q_i \left\{\sum_{\bfk\neq{\bf 0}}	g_{\mathbf{k}}^l(z_i,z) \cos\left[\mathbf{k}\cdot(\mathbf{r}-\mathbf{r}_i)\right]+
	\frac{e^{\frac{1}{2}k_{TF}(L+2z)}(2\varepssol +k_{TF}(L-2z_i))}{2\epsilon_0\ktf(2\varepssol +\ktf L)}\right\} 	 & -\infty<z<-\frac{L}{2}  \\
	\displaystyle \frac{1}{L_xL_y}\sum_{i}^{N_{\rm ions}}q_i \left\{ \sum_{\bfk\neq{\bf 0}} g_{\mathbf{k}}^s(z_i,z)\cos\left[\mathbf{k}\cdot(\mathbf{r}-\mathbf{r}_i)\right]+ 				\frac{[2\varepssol +k_{TF}(L+2z)][2\varepssol +k_{TF}(L-2z_i)]}{4\epsilon_0\ktf(2\varepssol +\ktf L)} \right\} 
	& -\frac{L}{2}<z<z_i \\
	\displaystyle \frac{1}{L_xL_y}\sum_{i}^{N_{\rm ions}}q_i \left\{ \sum_{\bfk\neq{\bf 0}} g_{\mathbf{k}}^s(z_i,z)\cos\left[\mathbf{k}\cdot(\mathbf{r}-\mathbf{r}_i)\right]+ 				\frac{[2\varepssol +k_{TF}(L-2z)][2\varepssol +k_{TF}(L+2z_i)]}{4\epsilon_0\ktf(2\varepssol +\ktf L)} \right\} 
	& z_i<z<\frac{L}{2} \\
	\displaystyle \frac{1}{L_xL_y}\sum_{i}^{N_{\rm ions}}q_i\left\{ \sum_{\bfk\neq{\bf 0}}	g_{\mathbf{k}}^r(z_i,z) \cos\left[\mathbf{k}\cdot(\mathbf{r}-\mathbf{r}_i)\right]+
	\frac{e^{\frac{1}{2}k_{TF}(L-2z)}(2\varepssol +k_{TF}(L+2z_i))}{2\epsilon_0\ktf(2\varepssol +\ktf L)} \right\}& \frac{L}{2} <z<+\infty\\
	\end{cases}
\end{eqnarray}
\end{widetext}
where the sums run over non-zero wave vectors corresponding to the periodic boundary conditions in the $x$ and $y$ directions along the surface, $\mathbf{k}=(\frac{2\pi m_x}{L_x},\frac{2\pi m_y}{L_y})$, and
for the left electrode, inter-electrode and right electrode regions:
\begin{widetext}
\begin{align}
\label{eq:phi:phiionsmodesl}
	g_{\mathbf{k}}^l(z_i,z) = & \displaystyle \frac{e^{\frac{1}{2}(\chi_{TF}(L+2z)-k(L+2z_i))}(e^{k(2z_i-L)}(\varepssol k-\chi_{TF})+(\varepssol k+\chi_{TF}))}{\varepsilon_0	\left[(1-e^{-2kL})(\varepssol ^2k^2+^2\chi_{TF}^2)+2(1+e^{-2kL})\varepssol k\chi_{TF}\right]}
	\\
\label{eq:phi:phiionsmodess}
	g_{\mathbf{k}}^s(z_i,z)=& \displaystyle  \frac{\left(e^{-k(z+z_i+L)}+e^{k(z+z_i-L)}\right)\left( \varepssol ^2  k^2 -\chi_{TF}^2\right)+e^{k(|z-z_i|-2L)}\left( \varepssol   k-\chi_{TF}\right)^2 + e^{-k|z_j-z_i|}\left( \varepssol   k+\chi_{TF}\right)^2}{2\varepsilon_0\varepssol k \left[(1-e^{-2kL})(\varepssol ^2k^2+^2\chi_{TF}^2)+2(1+e^{-2kL})\varepssol k\chi_{TF}\right]} 
    \\
\label{eq:phi:phiionsmodesr}
	g_{\mathbf{k}}^s(z_i,z) = & \displaystyle \frac{e^{\frac{1}{2}(\chi_{TF}(L-2z)-k(L-2z_i))}(e^{-k(2z_i+L)}(\varepssol k-\chi_{TF})+(\varepssol k+\chi_{TF}))}{\varepsilon_0 \left[(1-e^{-2kL})(\varepssol ^2k^2+^2\chi_{TF}^2)+2(1+e^{-2kL})\varepssol k\chi_{TF}\right]} \; .
\end{align}
\end{widetext}
In practice, in order to compute the effective many-body potential one only requires the potential experienced by the ions, \textit{i.e.} $g_{\mathbf{k}}^s(z_i,z)$ given in Eq.~\ref{eq:phi:phiionsmodess}. The terms outside the sums over modes in Eq.~\ref{eq:phi:phiions} correspond to $\bfk={\bf 0}$ and can be obtained as the $k\to0$ limits of the coefficients in Eq.~\ref{eq:phi:phiions}, so that one could in fact write the sum over all modes including $\bfk={\bf 0}$ for compactness. However, it is convenient to write this term separately to consider the limit $L\to\infty$, which is used to compute the energy $U_{\rm Coul}^{\rm ions}$ (see Eq.~\ref{eq:UCoulIonSepar}). In this limit, for ${\bf k}\neq0$ one has
\begin{align}
\label{eq:phi:phiionsmodesLinf}
	g_{\mathbf{k}}^\infty(z_i,z) &\equiv \lim_{L\to\infty} g_{\mathbf{k}}^s(z_i,z) = \displaystyle  \frac{e^{-k|z-z_i|}}{2\epsilon_0\varepssol k} \; ,
\end{align}
which corresponds to the 2D Fourier expansion of the Green's function for bulk ions in an implicit solvent, $1/4\varepsilon_0\varepssol|\bfr-\bfr_i|$. In the same limit, the ${\bf k}=0$ term in $\phi_{\rm ions}(\bfr)$ diverges but the corresponding $\mathcal{O}(L^2)$ and $\mathcal{O}(L)$ terms in the energy $\sum_i q_i \phi_{\rm ions}(\bfr_i)$ vanish for any finite $L$ due to the electroneutrality of the electrolyte, so that the limit $L\to\infty$ of the energy is well defined (see Eq.~\ref{eq:deltaUCoulIons}).


\bibliographystyle{aipnum4-1}
\bibliography{references}

\end{document}